\documentclass[aps,a4paper,prl,twocolumn,reprint,superscriptaddress,showpacs,prl]{revtex4-2}

\usepackage{graphicx}
\usepackage[english]{babel}
\usepackage{amsmath}
\usepackage{amsfonts}
\usepackage{amssymb}
\usepackage{xcolor}
\usepackage[normalem]{ulem}
\usepackage{tabularx}
\usepackage{comment}

\usepackage[colorlinks=true,linkcolor=blue, citecolor=blue, urlcolor=blue]{hyperref}

\begin{document}

\title{Orbital Topology of Chiral Crystals for Orbitronics}

\author {Kenta~Hagiwara}
\email{k-hagiwara@ims.ac.jp}
\affiliation {Peter Gr\"{u}nberg Institut (PGI-6), Forschungszentrum J\"{u}lich, J\"{u}lich 52425, Germany}
\affiliation {Faculty of Physics, University of Duisburg-Essen, Duisburg 47057, Germany}

\author {Ying-Jiun~Chen}
\email{yi.chen@fz-juelich.de}
\affiliation {Ernst Ruska-Centre for Microscopy and Spectroscopy with Electrons and Peter Gr\"{u}nberg Institute, Forschungszentrum J\"{u}lich, 52425 J\"{u}lich, Germany}

\author {Dongwook Go}
\email{dongo@uni-mainz.de}
\affiliation{Institute of Physics, Johannes Gutenberg University Mainz, 55099 Mainz, Germany}

\author {Xin~Liang~Tan}
\affiliation {Peter Gr\"{u}nberg Institut (PGI-6), Forschungszentrum J\"{u}lich, J\"{u}lich 52425, Germany}
\affiliation {Faculty of Physics, University of Duisburg-Essen, Duisburg 47057, Germany}

\author {Sergii~Grytsiuk}
\affiliation {Peter Gr\"{u}nberg Institut (PGI-1), Forschungszentrum J\"{u}lich and JARA, J\"{u}lich 52425, Germany}

\author{Kui-Hon~Ou~Yang}
\affiliation {Department of Physics, National Taiwan University, Taipei 10617, Taiwan}

\author{Guo-Jiun~Shu}
\affiliation {Department of Materials and Mineral Resources Engineering, National
Taipei University of Technology, Taipei 10608, Taiwan}

\author{Jing~Chien}
\affiliation {Department of Physics, National Taiwan University, Taipei 10617, Taiwan}
 
\author{Yi-Hsin~Shen}
\affiliation {Department of Physics, National Taiwan University, Taipei 10617, Taiwan}

\author{Xiang-Lin~Huang}
\affiliation {Department of Materials and Mineral Resources Engineering, National
Taipei University of Technology, Taipei 10608, Taiwan}

\author{Fang-Cheng~Chou}
\affiliation {Center of Condensed Matter Sciences, National Taiwan University, Taipei 106, Taiwan}

\author{Iulia~Cojocariu}
\altaffiliation[Present address: ] {Elettra-Sincrotrone Trieste S.C.p.A, Basovizza S.S. 14, Km 163.5, Trieste 34149, Italy, and Physics Department, University of Trieste, 34127 Trieste, Italy}
\affiliation {Peter Gr\"{u}nberg Institut (PGI-6), Forschungszentrum J\"{u}lich, J\"{u}lich 52425, Germany}

\author {Vitaliy~Feyer}
\affiliation {Peter Gr\"{u}nberg Institut (PGI-6), Forschungszentrum J\"{u}lich, J\"{u}lich 52425, Germany}
\affiliation {Faculty of Physics, University of Duisburg-Essen, Duisburg 47057, Germany}

\author{Minn-Tsong~Lin}
\affiliation {Department of Physics, National Taiwan University, Taipei 10617, Taiwan}
\affiliation{Institute of Atomic and Molecular Sciences, Academia Sinica, Taipei 10617, Taiwan}
\affiliation{Research Center for Applied Sciences, Academia Sinica, Taipei 11529, Taiwan}

\author {Stefan~Bl\"{u}gel}
\affiliation {Peter Gr\"{u}nberg Institut (PGI-1), Forschungszentrum J\"{u}lich and JARA, J\"{u}lich 52425, Germany}

\author {Claus~Michael~Schneider}
\affiliation {Peter Gr\"{u}nberg Institut (PGI-6), Forschungszentrum J\"{u}lich, J\"{u}lich 52425, Germany}
\affiliation {Faculty of Physics, University of Duisburg-Essen, Duisburg 47057, Germany}
\affiliation{Department of Physics, University of California Davis, Davis CA 95616, USA}

\author {Yuriy Mokrousov}
\affiliation {Peter Gr\"{u}nberg Institut (PGI-1), Forschungszentrum J\"{u}lich and JARA, J\"{u}lich 52425, Germany}
\affiliation{Institute of Physics, Johannes Gutenberg University Mainz, 55099 Mainz, Germany}

\author {Christian~Tusche}
\email{c.tusche@fz-juelich.de}
\affiliation {Peter Gr\"{u}nberg Institut (PGI-6), Forschungszentrum J\"{u}lich, J\"{u}lich 52425, Germany}
\affiliation {Faculty of Physics, University of Duisburg-Essen, Duisburg 47057, Germany}

\begin{abstract}
Chirality is ubiquitous in nature and manifests in a wide range of phenomena including chemical reactions,
biological processes, and quantum transport of electrons. In quantum materials, the chirality of fermions,
given by the relative directions between the electron spin and momentum, is connected to the band topology
of electronic states. Here, we show that in structurally chiral materials like CoSi, the orbital angular
momentum (OAM) serves as the main driver of a nontrivial band topology in this new class of unconventional topological semimetals, even when spin-orbit coupling is negligible.
A nontrivial orbital-momentum locking of multifold chiral fermions in the bulk
leads to a pronounced OAM texture of the helicoid Fermi arcs at the surface.
Our findings highlight the pivotal role of the orbital
degree of freedom for the chirality and topology of electron states, in general, and pave the way towards the application of topological chiral semimetals in orbitronic devices.

\end{abstract}

    \textcolor[rgb]{0.00,0.07,1.00}{}

\maketitle

\noindent
Chirality is ubiquitous in nature and manifested in a wide range of materials and phenomena, from the molecular biology of all organisms to the physical principles governing particle physics. Chiral structures come in pairs known as enantiomers, which are mirror images of each other and cannot be superimposed by rotation and translation. Opposite enantiomers may have completely different chemical/physical properties and respond differently to external stimuli. While many biomolecules that are essential to life exist as single enantiomers, 
utilizing enantiomeric pairs of quantum materials offers exciting opportunities for tailoring materials with specific functionality.
Chiral materials are fascinating due to their representation of the ultimate form of broken symmetry, which is distinct from other forms of symmetry breaking. Recent advancements highlight the potential of chiral materials in optoelectronics and spintronics, demonstrating applications like chiral-induced spin selectivity~\cite{na19}, circularly polarized beamsplitting~\cite{tur13}, and topological transport properties~\cite{ch18}.

The notion of chirality is also manifested in the low-energy excitation of electrons in a material. Topological Weyl semimetals host the analog of relativistic particles with a specific chirality (defined as the relative directions between the spin and momentum) in high energy physics, so-called Weyl fermions~\cite{Armitage2018, Lv2021}. The chirality of fermions is connected to the topology in the electronic structure through one of the topological numbers, the Chern number $C$.
A non-zero Chern number leads to the formation of a non-trivial Fermi arc at the surface by connecting a pair of Weyl points with opposite chirality, as the hallmark of a Weyl semimetal.
Generally, symmetry is decisive for the global topological properties in a material.
As outlined in Tab. \ref{table}, as more crystal symmetries are broken, i.e., the symmetries of the respective crystal structure are lower, a higher Chern number $C$ can be formed.
In non-magnetic materials, preserved space-inversion symmetry can lead to a Dirac semimetal with $C=0$.
Broken space-inversion symmetry can induce chirality in the electronic structure and give rise to a Weyl semimetal with $C=\pm1$ despite a non-chiral crystal structure.
Chiral-structured crystals have no space-inversion, mirror as well as other roto-inversion symmetries, and can behave more complex with $C=\pm2$ or $C=\pm4$~\cite{Chang2017}.
Moreover, chiral topological semimetals with $C=\pm2$ and $C=\pm4$ belong to a ``unconventional'' class, where unconventional fermions emerge which have no analogs in high-energy physics \cite{Bradlyn2016}. These low-energy excitations are denoted by multifold chiral fermions~{\cite{ch18}}.

As schematically illustrated in Tab. \ref{table} for the CoSi-family topological chiral semimetal, high Chern numbers $C=\pm 2$ of multifold chiral fermions are found at the $\Gamma$ and $\mathrm{R}$ points. As they are ``maximally separated'' in the momentum space, exceptionally long Fermi arcs exist at the surface, distinct from conventional Weyl semimetals. Remarkably, the emergence of the multifold chiral fermion in chiral crystals does not require spin-orbit coupling (SOC), but instead is driven by the structural chirality.
The additional consideration of SOC, albeit small, leads to an even larger $C=\pm4$, and spin-split
Fermi arcs~\cite{Takane2019, Rao2019, Sanchez2019}.

\renewcommand{\arraystretch}{1.3}

\begin{table}[t!]
\fontsize{9}{10} \selectfont
\linespread{1.0}
\begin{tabular}{|p{1.55cm}|p{1.7cm}|p{2.2cm}|p{2.5cm}|} 
    \hline \hline Semimetals & \multicolumn{1}{c|}{Dirac} & \multicolumn{1}{c|}{Weyl} & \multicolumn{1}{c|}{Chiral} \\ \hline \hline
    Crystal \newline structure & Preserved space-inversion symmetry & Broken~space-inversion symmetry & Chiral structure (lack of inversion, mirror or other roto-inversion symmetries)  \\ \hline
    Symmetry & \multicolumn{1}{c|}{High} &  & \multicolumn{1}{c|}{Low}  \\ \hline
    Chern number & 
    \begin{minipage}{0.2cm}
      \centering
      \vspace{0.2cm}
      \scalebox{0.35}{\includegraphics{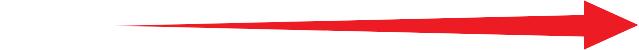}}
    \end{minipage} & &
    \\
    & \multicolumn{1}{c|}{0} & \multicolumn{1}{c|}{$\pm1$} & \multicolumn{1}{c|}{$\pm2~(\pm4)$} \\ \hline
    Band \newline structure &
    \begin{minipage}{2cm}
      \centering
      \scalebox{0.3}{\includegraphics{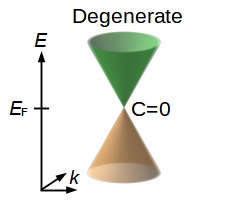}}
    \end{minipage} &
    \begin{minipage}{2cm}
      \centering
      \scalebox{0.3}{\includegraphics{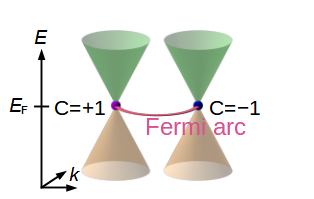}}
    \end{minipage} &
    \begin{minipage}{2cm}
      \centering
      \scalebox{0.4}{\includegraphics{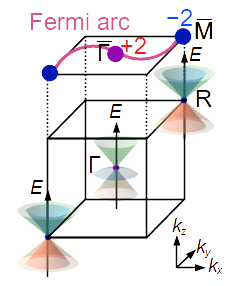}}
    \end{minipage} \\ \hline 
    Materials & Na$_3$Bi, Cd$_3$As$_2$, NiTe$_{2}$ & (Ta,~Nb)(As,~P) Co$_3$Sn$_2$S$_2$, Mn$_3$Sn, Mn$_3$Ge, Mo$_{x}$W$_{1-x}$Te$_{2}$ & CoSi,\,RhSi,\,PdGa, PtGa \\ \hline \hline
\end{tabular}
\caption{Overview of topological semimetals depending on the symmetry of crystal structures.}
\label{table} 
\end{table}

In this work, we unveil the nature of the interplay in the topological chiral semimetal CoSi by means of advanced momentum microscopy, which can measure both bulk and surface electronic states with high sensitivity. We utilize the circular dichroism (CD), which serves as a probe for the orbital degree of freedom of the electronic states~\cite{Min2019,Uenzelmann2021,Park2012,Park2012a,Schueler2020}. Our experimental results show a distinct chirality of the orbital angular momentum (OAM) texture in CoSi, which undergoes a sign change between chiral crystals of opposite handedness. The nontrivial topology of the multifold chiral fermions originates from the orbital chirality in the bulk, and is evidenced by a pronounced OAM polarization of the topological helicoid Fermi arcs. Our first-principles calculations confirm that the OAM serves as the key link among the crystal chirality, electronic chirality, and topology.

\begin{figure*}[]
\centering
\includegraphics[width=0.83\textwidth]{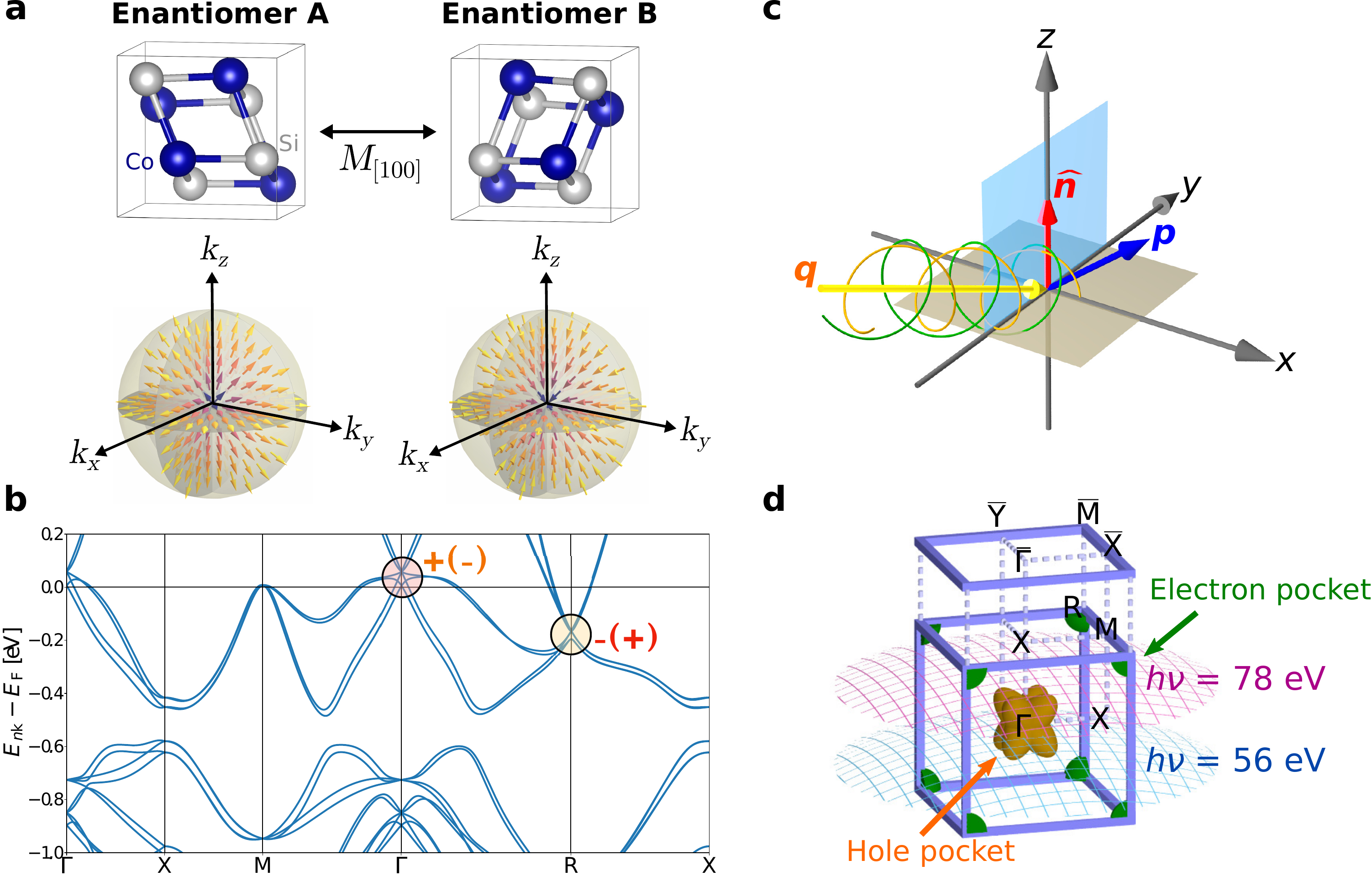}
\caption{
{\bf Topological chiral semimetal CoSi and schematics of the experiment.}
{\bf a,} The crystal structures of enantiomers A (left) and B (right) are related by the mirror reflection along [100] axis. Below, the $\mathbf{k}$-space OAM texture of a multifold chiral fermion is schematically illustrated. For enantiomers A and B, the orbital chirality $\mathbf{L}\cdot\mathbf{k}$ are the opposite. {\bf b,} Band structure of CoSi. Bulk $\Gamma$ and $R$ points with high Chern numbers are indicated by circles,. These charges have opposite signs for enantiomers A and B due to the opposite orbital chirality.
\textbf{c,} Experimental photoemission geometry.
The photon incidence vector (\textbf{q}) lies in the $yz$ plane at an angle of 65$^\circ$ with respect to the sample surface normal ($\mathbf{\hat{n}}$) and is aligned along the $\overline{\text{Y}}-\overline{\Gamma}-\overline{\text{Y}}$ direction of the surface BZ. Emitted photoelectrons (\textbf{p}) are collected in the full solid angle above the sample surface ($xy$-plane) by the momentum microscope.
\textbf{d,} Schematic Fermi surface of CoSi in the bulk BZ with corresponding spherical sections measured at different photon energies. 
}
\label{fig:overview}
\end{figure*}

\ \\ 
\noindent {\bf The role of the OAM in the interplay among crystal chirality, electronic chirality and topology}

\noindent 
The crystal structure of CoSi, which is one of the prototypical chiral topological semimetals, is shown in Fig.~\ref{fig:overview}{\bf a}. The two enantiomer pairs A and B are related via a mirror reflection with respect to the [100] axis. Below, we show the schematics of the OAM textures in $\mathbf{k}$-space of a multifold chiral fermion, where the arrows represent the OAM. The radial and monopole-like OAM texture has been previously predicted~\cite{Yang2023}. The band structure of bulk CoSi is shown in Fig.~\ref{fig:overview}{\bf b}. The multifold chiral fermions are found at $\Gamma$ and $\mathrm{R}$, which are highlighted in circles. Our first-principles calculations clearly demonstrate the OAM chirality at these points (Supplementary Information).

We define the OAM chirality of electronic states as the sign of $\mathbf{L}\cdot\mathbf{k}$ in $\mathbf{k}$-space. The monopole charges associated with the OAM chirality are opposite for enantiomers A and B. This can be shown by the mirror reflection $M_{\mathrm{[100]}}$, which interchanges enantiomers A and B and transforms the momentum and OAM by $(k_x, k_y, k_z) \rightarrow (-k_x, k_y, k_z)$ and $(L_x, L_y, L_z)\rightarrow (L_x, -L_y, -L_z)$. Thus, positive chirality $\mathbf{L} \cdot {\mathbf{k}}>0$, in enantiomer A, implies negative chirality $\mathbf{L}\cdot {\mathbf{k}}<0$ in enantiomer B. We emphasize that although the spin also exhibits chirality, it \emph{derives} from the OAM chirality via SOC. The spin splitting is significantly smaller than the OAM splitting in the band structure, and disappears when the SOC is neglected (Supplementary Information). Therefore, the OAM is the key link between the chirality of crystal structures and the chirality of the OAM texture in $\mathbf{k}$-space.

In particular, the Berry curvature in the topological chiral semimetal CoSi is of \emph{orbital origin}, without requiring SOC for its emergence. Thus, the opposite signs of the OAM chirality can be directly translated to opposite topological charges and the Berry curvature fields in $\mathbf{k}$-space. One of the striking manifestations of the orbital origin of the topological charges in the bulk is the strong OAM polarization of the Fermi arcs, which we confirm in our work. We emphasize that in bulk CoSi, enantiomers A and B have exactly the same band structure. However, the opposite OAM chirality as well as the topological charges suggest that the Fermi arcs at the surface exhibit differently shaped profiles for enantiomers A and B. Thus, measuring Fermi arcs can be a way to distinguish the two enantiomers in an experiment. We also note that the rotation sense of the \emph{helicoid} Fermi arcs, a helix-like feature in the energy domain, also differs in enantiomers A and B due to the opposite OAM chirality of the multifold chiral fermions in the bulk.

\begin{figure*}[t!]
	\centering
	\includegraphics[width=0.83\textwidth]{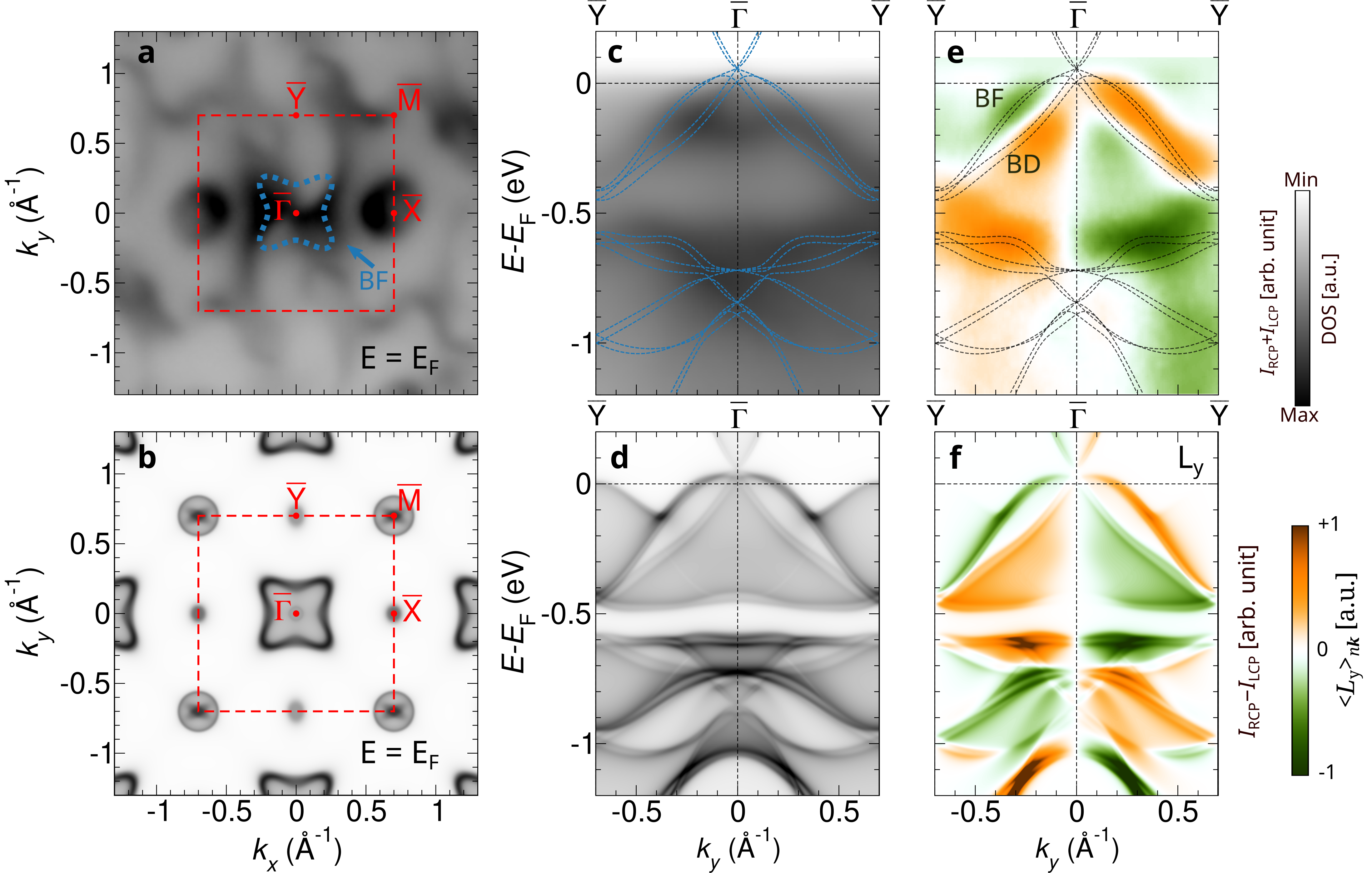}
	\caption{
 {\bf OAM fingerprint of multifold chiral fermions in the bulk.}
  {\bf a-d,} Measured photoemission intensity ({\bf a,c}) and calculated spectral weight ({\bf b,d}) projected on the surface BZ at the Fermi energy and along the $\overline{\text{Y}}-\overline{\Gamma}-\overline{\text{Y}}$ path as a function of the energy. {\bf e,} Measured CD map and {\bf f,} calculated $L_y$-projected spectral weight. Bulk flat (BF) and bulk Dirac (BD) bands are indicated in {\bf a} and {\bf e}. In {\bf c} and {\bf e}, the calculated bulk band structure is shown by dashed lines.
		}
	\label{fig:bulk}
\end{figure*}

\ \\ 
\noindent {\bf Detection of the OAM by momentum microscope}

\noindent
To experimentally measure the electronic band structure and the OAM texture, we employed a momentum-resolved photoemission microscope setup at the NanoESCA beamline of the Elettra synchrotron in Trieste (Italy) \cite{Wiemann2011}.
In a momentum microscope, photoelectrons emitted into the complete solid angle above the sample surface can be collected simultaneously, such that the two-dimensional (2D) photoelectron distribution of the in-plane crystal momentum ($k_x, k_y$) is be obtained in a single measurement, covering a wide region over the whole Brillouin zone (BZ) \cite{tus15,Tusche2019}. This is one of the major advantages of the momentum microscope, distinct from other techniques such as angle-resolved photoemission spectroscopy. The photoemission experiments were performed in the geometry schematically shown in Fig.~\ref{fig:overview}{\bf c}.
The plane of incidence coincides with the $\overline{\Gamma}-\overline{\text{Y}}$ direction of the surface BZ., which thus represents a distinguished plane of high symmetry for the experimental geometry.

To confirm whether the structural chirality reflects in the electronic structure via orbital-momentum locking, we carried out momentum-resolved photoemission experiments using left- and right-circularly polarized light (LCP and RCP).
The examination of CD in the angular distribution (CDAD) extends beyond the band structure mapping, providing a more comprehensive exploration of the momentum-resolved Bloch wave function, including orbitals~\cite{Park2012,Park2012a,Uenzelmann2021}, Berry curvature~\cite{Schueler2020,Liu2011a,Uenzelmann2021}, and topological invariants~\cite{Hagiwara2022,Uenzelmann2021}. Moreover, this method allows us to distinguish the handedness of the chiral crystals and their surfaces~\cite{fe22}. The OAM contribution in the CDAD~\cite{sc22} can be intuitively understood by considering both the helicity of light and the self-rotation of the Bloch states~\cite{Schueler2020,Park2012a}. In nonmagnetic materials, the combination of time-reversal and spacial inversion symmetries is a strong condition of orbital quenching, and dictates that OAM is absent for all states~\cite{Go2021}. The broken symmetry in a chiral crystal,
however, can give rise to the presence of OAM at each $\mathbf{k}$-point.

We note that circularly polarized light couples not only to an intrinsic chiral system, but also to a chirality introduced by the experimental setup itself~\cite{Schoenhense1990}. That is, the incoming photon vector \textit{\textbf{q}}, the photoelectron momentum \textit{\textbf{p}}, and the surface normal $\hat{\textit{\textbf{n}}}$ define a handed coordinate system.
The CDAD contribution arising from the experimental geometry vanishes on the $k_x=0$ line ($\overline{\Gamma}-\overline{\text{Y}}$ direction), where all three vectors, \textit{\textbf{q}}, \textit{\textbf{p}}, and $\hat{\textit{\textbf{n}}}$ lie in the same plane. Therefore, the CDAD along the $\overline{\Gamma}-\overline{\text{Y}}$ direction reflects the intrinsic chirality of the electronic states.

By using different photon energies, which select different $k_z$ values (see Fig.~\ref{fig:overview}{\bf d})~\cite{Rao2019,ta23,tu18}, we predominantly examine either multifold chiral fermion states in the bulk or helicoid Fermi arc states at the surface. At $\hbar \nu = 56\ \mathrm{eV}$, we probe a section through the center of the hole pocket at the $\Gamma$ point of the bulk BZ, where the presence of multifold chiral fermions is anticipated. At $\hbar \nu = 78\ \mathrm{eV}$, the $k_z$ cut is positioned between the hole and electron pockets of the Fermi surfaces in the bulk BZ.

\ \\
\textbf{OAM chirality of multifold chiral fermions}

\begin{figure}[]
	\centering
	\includegraphics[width=0.47\textwidth]{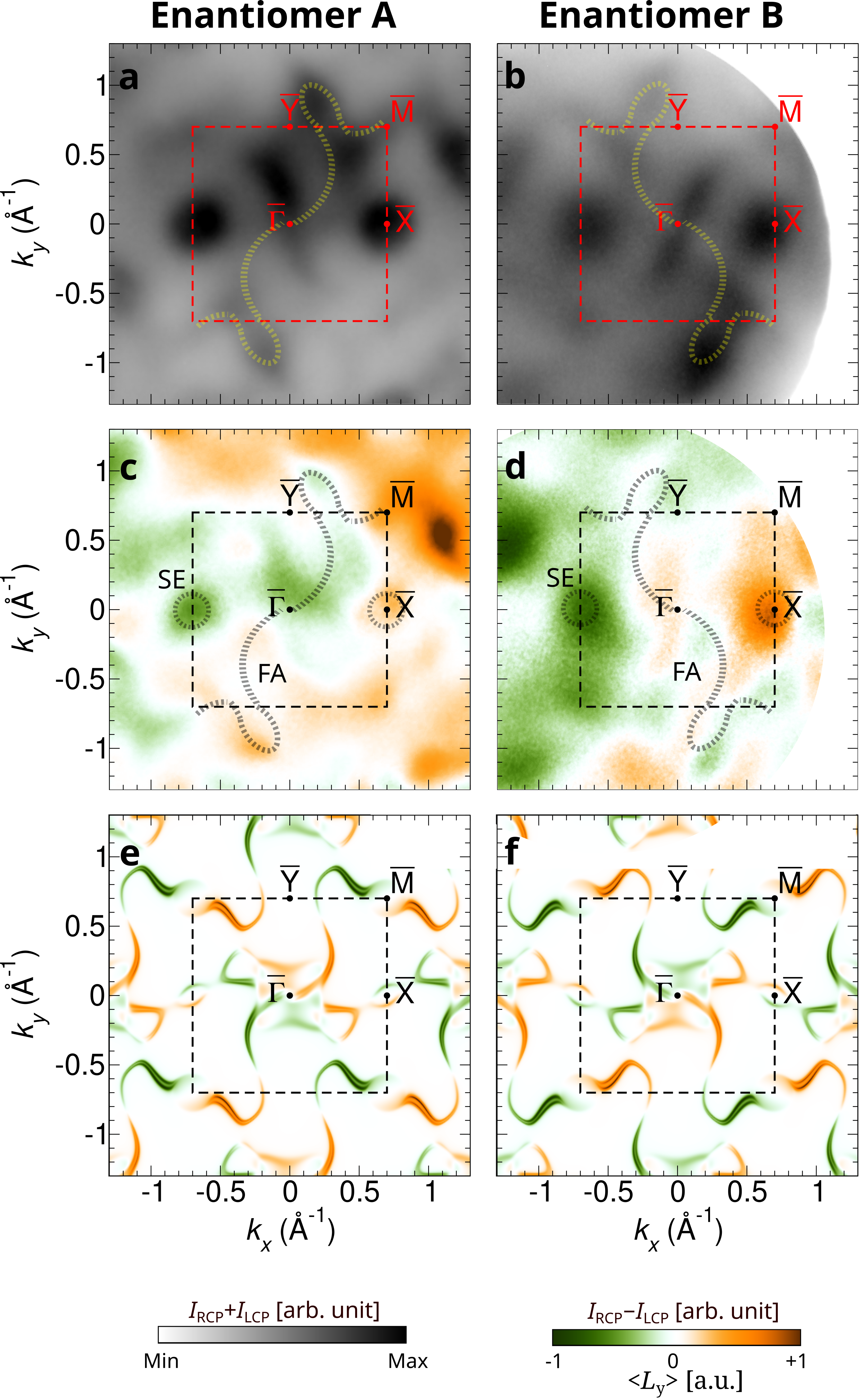}
	\caption{{\bf Chirality-driven orbital texture of helicoid topological Fermi arcs.}
		\textbf{a,b,} The sum of intensities of photoemission momentum maps measured by LCP and RCP light at the Fermi energy for enantiomers A and B. {\bf c,d,} The measured CD maps for enantiomers A and B. {\bf e,f,} Calculated spectral weight at a surface with $L_y$ projection, for enantiomers A and B. We denote the Fermi arcs and surface electrons by FA and SE, respectively.
	}
	\label{fig:surface}
\end{figure}

\noindent
Multifold chiral fermions at $\Gamma$ are found in the bulk band structure of CoSi (Fig.~\ref{fig:overview}{\bf b}), which can be accessed in our photoemission experiment with a photon energy of $56\ \mathrm{eV}$. Here, we focus on enantiomer A. Figure~\ref{fig:bulk}{\bf a} shows the 2D photoemission intensity map on the surface BZ at the Fermi energy. The intensity is measured by summing up the intensities recorded with LCP and RCP light ($I_\mathrm {Total}=I_\mathrm {LCP}+I_\mathrm {RCP}$). As marked by the blue dashed line, bulk flat (BF) bands of the chiral fermions are observed at E$_\mathrm{F}$. This feature is also found in the calculated spectral weight shown in Fig.~\ref{fig:bulk}{\bf b}. The four-fold rotation-symmetric Fermi surface is clearly visible because of the presence of a screw rotation symmetry ($90^\circ$ rotation followed by a spatial translation). We note that in Fig.~\ref{fig:bulk}{\bf a}, we find features that violate the four-fold rotation symmetry due to surface states including the topological helicoid arcs. Here, we focus mainly on the multifold chiral fermions in the bulk, while features of the surface state are revisited in Fig.~\ref{fig:surface}. In Fig.~\ref{fig:bulk}{\bf c}, photoemission intensities along the $\overline{\text{Y}}-\overline{\Gamma}-\overline{\text{Y}}$ direction are shown as a function of the energy, together with the calculated bulk band structure (dashed lines). In comparison with the computed spectral weight projected on the surface BZ (Fig.~\ref{fig:bulk}{\bf d}), a good agreement between theory and experiment is obtained.

The OAM of the multifold chiral fermions can be assessed by the CD ($I_\mathrm {CD}=I_\mathrm {LCP}-I_\mathrm {RCP}$). As shown in Fig.~\ref{fig:bulk}{\bf e}, the pronounced CD signal along the $\overline{\text{Y}}-\overline{\Gamma}-\overline{\text{Y}}$ direction signifies the OAM,
suggesting substantial contributions to the Berry curvature in momentum space. Notably, substantial CD signals within the optical plane reveal a significant OAM $L_y$, parallel to the momentum $k_y$. The sign of the CD signal is an odd function of $k_y$, being consistent with the monopole-like radial texture of the OAM, giving rise to the chirality ($\mathbf{k}\cdot\mathbf{L}$) of the multifold chiral fermions. We find that the BF and bulk Dirac (BD) bands exhibit the opposite OAM chirality. We corroborate these findings by theory as well. Figure~\ref{fig:bulk}{\bf f} shows the calculated $L_y$-projected spectral weight on the surface BZ, which is in excellent agreement with the CD measured in our experiment (Fig.~\ref{fig:bulk}{\bf e}). The weights for the other components of the OAM are shown in the Supplementary Information. We emphasize that the OAM polarization of the helicoid Fermi arc is of nonrelativistic origin, which depends on the crystal chirality. The spin texture can arise with the aid of SOC, but since SOC is much weaker compared to the crystal-field potential in this material, the spin splittings are much smaller, i.e. spin up and down bands are nearly degenerate (Supplementary Information).

\ \\ 
\noindent {\bf OAM texture of helicoid topological Fermi arcs}

\noindent At a photon energy of $78\ \mathrm{eV}$, the photoemission signal from the bulk electron and hole pockets can be reduced (Fig.~\ref{fig:overview}{\bf d}), which allows us to better analyze the surface states. The total intensity map ($I_\mathrm {Tot}=I_\mathrm {LCP}+I_\mathrm {RCP}$) over the surface BZ at the Fermi energy is shown in Figs.~\ref{fig:surface}{\bf a} and \ref{fig:surface}{\bf b}, for enantiomers A and B, respectively. We clearly observe the absence of the four-fold rotation symmetry. This is because the screw rotation symmetry is missing at the surface. Importantly, the ``direction'' of symmetry breaking is opposite between enantiomers A and B. Thus, although the bulk spectra are identical for both enantiomers, examining the surface allows to distinguish them.

One of the most distinctive spectral features at the surface of CoSi is the emergence of helicoid topological Fermi arcs. They originate from the difference of the Chern numbers of the multifold chiral fermions at the ${\Gamma}$ and ${\mathrm{R}}$ points in the bulk (Fig.~\ref{fig:overview}{\bf b}). Thus, the presence of the Fermi arcs is guaranteed by the bulk-boundary correspondence principle in the surface BZ between $\overline{\Gamma}$ and $\overline{\mathrm{M}}$. These points are maximally separated in momentum space (Tab.~\ref{table}). In general, how the chiral charges are connected by Fermi arcs depends on details of the microscopic interactions at the surface. Interestingly, this connection is different for enantiomers A and B. As indicated by dashed lines in Fig.~\ref{fig:surface}{\bf a} and \ref{fig:surface}{\bf b}, the Fermi arcs of enantiomers A and B connect between $\Gamma$ and $(+0.7, +0.7)$\,\AA$^{-1}$
and between $\Gamma$ and $(-0.7, +0.7)$\,\AA$^{-1}$
which reflects the breaking of the screw rotation symmetry at the surface. Despite this difference, the Fermi arcs of enantiomers A and B are related by mirror reflection along the $[100]$ axis.

Another striking difference between the Fermi arcs of enantiomer A and B is their OAM polarization. The main reason for this is due to different signs of the topological charges for enantiomers A and B, which ultimately originates from the difference in the crystal chirality. In Fig.~\ref{fig:surface}{\bf c} and \ref{fig:surface}{\bf d}, the CD ($I_\mathrm {CD}=I_\mathrm {LCP}-I_\mathrm {RCP}$) is shown over the surface BZ, measured on the surfaces of enantiomers A and B, respectively. The results show that the Fermi arcs exhibit a pronounced OAM polarization, which shows a clear difference between enantiomers A and B, related by a mirror reflection along the $a$-axis.

To understand the difference of the Fermi arcs in enantiomers A and B, in Figs.~\ref{fig:surface}{\bf e} and \ref{fig:surface}{\bf f}, we show the calculated results of the $L_y$-projected spectral weights,
respectively. The surface spectrum and the OAM polarization are calculated by the surface Green's function technique for a semi-infinite geometry, based on the Wannier representation of the bulk periodic electronic structure. In general, we find that the overall shapes of the Fermi arcs agree with the experimental results (Figs.~\ref{fig:surface}{\bf a} and \ref{fig:surface}{\bf b}); the calculation clearly shows that the Fermi arcs connect between $\overline{\Gamma}$ and $\overline{\mathrm{M}}$,
and exhibit an opposite rotation sense in enantiomers A and B.

The calculated $L_y$ agrees well with the experimental result (Figs.~\ref{fig:surface}{\bf c} and \ref{fig:surface}{\bf d}) for the Fermi arcs. In particular, for the states near $\overline{\mathrm{M}}$, the sign change of $L_y$ is well reproduced at the Brillouin zone boundary. However, a quantitative comparison of $L_y$ is challanging for the states near $\overline{\Gamma}$. This can be understood by the large spectral weight of the bulk states: as schematically illustrated in Fig.~\ref{fig:overview}{\bf d}, the hole pocket near $\overline{\Gamma}$ is substantially larger than the electron pocket near $\overline{\mathrm{M}}$, which is also confirmed by our first-principles calculation (Supplementary Information). Furthermorre, we remark that in general the OAM polarization depends strongly on the electric polarization at a surface~\cite{Park2011, Go2017, Go2021a}. That is, the detailed sign and magnitude of the orbital polarization may significantly change depending on the condition at the surface, whose assumption in the theory generally differs from the condition in the experiment. For example, the atoms at a surface may undergo a structural relaxation, which is not considered in the calculations.

Nonetheless, the \emph{symmetry relation} between the OAM textures of enantiomers A and B, in terms of the mirror reflection along the $a$-axis, is robust in both theory and experiment. That is, $L_y (k_x, k_y)$ in enantiomer A is equal to $-L_y (-k_x, k_y)$. Meanwhile, we also remark that the SOC in this system is small, resulting in a spin splitting of the Fermi arcs that is far too small to be detected in the experiments (Supplementary Information). The OAM textures of the Fermi arcs are a direct consequence of the crystal-field potential, not the SOC.

Beside the helicoid Fermi arcs, also other surface states are observed:
a disk shaped state at the $\overline{\mathrm{X}}$ point forms a surface electron (SE) pocket
as indicated in Figs.~\ref{fig:surface}{\bf c} and \ref{fig:surface}{\bf d}. These states are of a non-topological origin, e.g., the orbital Rashba interaction at the surface. Thus, the OAM polarizations at $\overline{\mathrm{X}}$ are the same for both enantiomers A and B. Interestingly, this is not found in the semi-infinite slab calculation, but a fully consistent first-principles calculation of a finite slab reproduces the large spectral weight at $\overline{\mathrm{X}}$ (Supplementary Information). This suggests the SE states rely sensitively on the boundary condition at the surface, and thus are independent from the band topology in the bulk.

\ \\ 
\noindent{\bf Concluding remarks and outlook}

\noindent 
In this work, we have shown the crucial role of the OAM as the key link intertwining crystal chirality, electronic chirality and band topology in the topological chiral crystal CoSi by momentum microscopy experiments and first-principles calculations. In the bulk, the radial OAM texture is not only a characteristic feature of multifold chiral fermions but also the origin of a high Chern number. As the orbital degree of freedom directly interacts with the crystal-field potential, enantiomers A and B exhibit an opposite sign of the OAM texture, Berry curvature, and the Chern number of their electronic bands. The nontrivial topology of the multifold chiral fermions in the bulk manifests at the surface, where helicoid Fermi arcs connect the surface projections of the two bulk topological charges. We demonstrate the orbital origin of the helicoid Fermi arcs by observing a pronounced OAM texture, which depends on the crystal chirality.

Our work bridges the two separate but important areas of chiral chemistry/physics and topological materials. On the one hand, the notion of atomic and molecular orbitals is the natural language in chiral chemistry, as they couple directly to chiral structures. However, the orbitals have been regarded as mainly responsible for the stability of molecules and materials and not considered as the ``information'' or the degree of freedom. On the other hand, in the study of topological materials, the role of the orbital degree of freedom has been overlooked; historically most early findings have been associated with SOC and the spin-momentum locking. While this early development was driven by analogy with particle physics, as we pointed out, the multifold chiral fermion, which originates from the orbital degree of freedom, finds no analogy among elementary particles. Therefore, in the new interdisciplinary field of chiral physics in topological materials, it is encouraged to investigate the manifestations of the OAM further, e.g. in transport and optical phenomena, as well as reaction dynamics, in a unified framework. Promoting the OAM as a foundation for topological characterization is another pending challenge, which has to be addressed in the future. Chirality emerges as an analog of spin or magnetization for spin and charge derived topologies -- a new Ising-like variable for topological classification rooting in OAM of electronic states. In this context the OAM-powered chirality-sensitive Fermi arcs will be crucial for understanding linear and non-linear (magneto-)trasport orbitally-flavored phenomena arising at the boundaries between domains with different crystal chirality.

In this context, our findings may resonate with recent developments of ``orbitronics'', which aims to exploit the OAM or the orbital degree of freedom of electrons as an information carrier in next-generation electronic devices~\cite{Go2021}. Orbital-momentum locking present in both bulk and surface states
makes topological chiral crystals an intriguing material platform to discover novel quantum transport phenomena beyond the orbital Hall effect and the orbital Edelstein effect~\cite{Calavalle2022, Roy2022}. For example, the orbital texture of the helicoid Fermi arcs may be engineered for orbitronic devices and harnessed by converting between OAM and  the electron spin for spintronic devices. The interplay of structural chirality with electronic OAM at the surface of chiral metals presents other exciting possibilities for orbitronics realizations, such as, e.g., excitation of chiral phonons~\cite{Ishito2023, Ueda2023} by electrical currents and orbital relaxation, OAM-mediated structural chirality switching by circularly polarized light or effects of orbital filtering in magneto-resistance type of transport applications.

\section*{Methods}

\noindent 
{\bf Sample preparation}
\ \\ 
\noindent 
High-quality single crystals were grown by the modified Bridgeman method with the help of an optical floating-zone furnace as described in Ref.~\cite{OuYang2015, OuYang2015a} and polished on the (001) surface.
Since the Co and Si atoms are strongly bonded by multiple covalent bonds in three dimensions and the high symmetry surfaces are not cleavage planes, we performed Ar$^{+}$ ion sputtering and annealing in the preparation chamber of the NanoESCA beamline in order to get a clean sample surface instated of cleaving the sample.
Referring to the cleaning procedure of the transition-metal monosilicides FeSi \cite{Castro1997, Klein2009} and MnSi \cite{Kura2008}, we obtained the best condition for the \textit{in situ} surface preparation of CoSi: Ar$^{+}$ ion sputtering with an energy of 2\,keV and annealing at $T = 680^\circ$C.

\ \\ 
\noindent 
{\bf Momentum microscopy measurements}
\ \\ 
\noindent 
Momentum-resolved photoelectron spectroscopy experiments were carried out at the NanoESCA beamline~\cite{Wiemann2011} of the Elettra synchrotron in Trieste (Italy), using left- and right-circularly polarized light, without changing the overall experimental geometry. All measurements were performed while keeping the sample at a temperature of 130~K. Photoelectrons emitted into the complete solid angle above the sample surface were collected using a momentum microscope~\cite{tus15,tus19}. The momentum microscope directly forms an image of the distribution of photoelectrons as function of the lateral crystal momentum ($k_x$,$k_y$) that is recorded on an imaging detector~\cite{tus15,tus19}.

\ \\ 
\noindent 
{\bf First-principles calculations}
\ \\ 
\noindent 
The electronic band structure and the OAM texture are computed by the Wannier interpolation method from first-principles, which consists of three steps of calculation. First, we converge the charge density and potential by self-consistent density functional theory calculation. Here, we employ the FLEUR code~\cite{fleur}, in which the full-potential linearly augmented plane wave method~\cite{Wimmer1981} is implemented. The exchange and correlation effects are treated within the generalized gradient approximation by using the Perdew-Burke-Ernzerhof functional~\cite{Perdew1996}. The lattice constant of the simple cubic structure is set $8.38 a_0$, where $a_0$ is the Bohr radius, wherein the atom positions are listed in Tab.~\ref{tab:atoms}. We set the maximum of the harmonic expansion in the muffin-tin sphere $l_\mathrm{max}=12$ and the muffin-tin radii of both Co and Si atoms $R_\mathrm{MT}=2.13a_0$. The plane wave cutoff is set $K_\mathrm{max}=5.0a_0^{-1}$. We sample $\mathbf{k}$-points from the Monkhorst-Pack mesh of $16\times 16\times 16$. The SOC is considered in the fully relativistic manner by the second variation scheme.

\newcolumntype{M}[1]{>{\centering\arraybackslash}m{#1}}
\renewcommand{\arraystretch}{1.3}
\begin{table}[t!]
\fontsize{9}{10} \selectfont
\linespread{1.0}
\centering
\begin{tabular}{| M{2cm} | M{4cm} |}
\hline 
\hline Atom & Fractional coordinate \\ 
\hline 
\hline Co-1 & (0.1451, 0.1451, 0.1451) \\ 
\hline Co-2 & (0.3549, 0.8549, 0.6451) \\ 
\hline Co-3 & (0.6451, 0.3549, 0.8549) \\
\hline Co-4 & (0.8549, 0.6451, 0.3549) \\
\hline Si-1 & (0.1568, 0.3432, 0.6568) \\
\hline Si-2 & (0.3432, 0.6568, 0.1568) \\
\hline Si-3 & (0.6568, 0.1568, 0.3432) \\
\hline Si-4 & (0.8432, 0.8432, 0.8432) \\
\hline
\hline
\end{tabular}
\caption{The positions of Co and Si atoms in the fractional coordinate of the simple cubic unit cell.}
\label{tab:atoms} 
\end{table}

Second, we construct the maximally localized Wannier functions from the converged Kohn-Sham states from the first step. As the initial guess, we project the Kohn-Sham states onto $d_{xy}$, $d_{yz}$, $d_{zx}$, $d_{z^2}$, $d_{x^2-y^2}$ orbitals on Co sites and $s$, $p_x$, $p_y$, $p_z$ orbitals on Si sites. The spread of Wannier functions are iteratively minimized by using the WANNIER90 code~\cite{Pizzi2020}. For the disentanglement step, we set the frozen energy window such that its maximum is $2\ \mathrm{eV}$ above the Fermi energy. The Hamiltonian, position, spin, and OAM operators, which are initially evaluated in the Kohn-Sham basis, are transformed into the Wannier basis by using the interface between the FLEUR and WANNIER90 codes~\cite{Freimuth2008}. The OAM is evaluated within the atom-centered approximation~\cite{Hanke2016}.

Third, we use the home-built code ORBITRANS to evaluate the OAM expectation value of the bulk band structure as well as the OAM-projected spectral weight of the in the semi-infinite geometry. For the latter, we implement the surface Green's function method for the consideration of the semi-infinite slab by following the scheme in Ref.~\cite{Sancho1985}. For the Wannier interpolation along high-symmetry points, we use $300$ points, and for the plots on the surface BZ, we use a $200\times 200$ meshgrid.

\bibliography{references.bib}

\begin{thebibliography}{47}%
\makeatletter
\providecommand \@ifxundefined [1]{%
 \@ifx{#1\undefined}
}%
\providecommand \@ifnum [1]{%
 \ifnum #1\expandafter \@firstoftwo
 \else \expandafter \@secondoftwo
 \fi
}%
\providecommand \@ifx [1]{%
 \ifx #1\expandafter \@firstoftwo
 \else \expandafter \@secondoftwo
 \fi
}%
\providecommand \natexlab [1]{#1}%
\providecommand \enquote  [1]{``#1''}%
\providecommand \bibnamefont  [1]{#1}%
\providecommand \bibfnamefont [1]{#1}%
\providecommand \citenamefont [1]{#1}%
\providecommand \href@noop [0]{\@secondoftwo}%
\providecommand \href [0]{\begingroup \@sanitize@url \@href}%
\providecommand \@href[1]{\@@startlink{#1}\@@href}%
\providecommand \@@href[1]{\endgroup#1\@@endlink}%
\providecommand \@sanitize@url [0]{\catcode `\\12\catcode `\$12\catcode
  `\&12\catcode `\#12\catcode `\^12\catcode `\_12\catcode `\%12\relax}%
\providecommand \@@startlink[1]{}%
\providecommand \@@endlink[0]{}%
\providecommand \url  [0]{\begingroup\@sanitize@url \@url }%
\providecommand \@url [1]{\endgroup\@href {#1}{\urlprefix }}%
\providecommand \urlprefix  [0]{URL }%
\providecommand \Eprint [0]{\href }%
\providecommand \doibase [0]{http://dx.doi.org/}%
\providecommand \selectlanguage [0]{\@gobble}%
\providecommand \bibinfo  [0]{\@secondoftwo}%
\providecommand \bibfield  [0]{\@secondoftwo}%
\providecommand \translation [1]{[#1]}%
\providecommand \BibitemOpen [0]{}%
\providecommand \bibitemStop [0]{}%
\providecommand \bibitemNoStop [0]{.\EOS\space}%
\providecommand \EOS [0]{\spacefactor3000\relax}%
\providecommand \BibitemShut  [1]{\csname bibitem#1\endcsname}%
\let\auto@bib@innerbib\@empty
\bibitem [{\citenamefont {Naaman}, \citenamefont {Paltiel},\ and\ \citenamefont
  {Waldeck}(2019)}]{na19}%
  \BibitemOpen
  \bibfield  {author} {\bibinfo {author} {\bibfnamefont {R.}~\bibnamefont
  {Naaman}}, \bibinfo {author} {\bibfnamefont {Y.}~\bibnamefont {Paltiel}}, \
  and\ \bibinfo {author} {\bibfnamefont {D.~H.}\ \bibnamefont {Waldeck}},\
  }\bibfield  {title} {\enquote {\bibinfo {title} {Chiral molecules and the
  electron spin},}\ }\href@noop {} {\bibfield  {journal} {\bibinfo  {journal}
  {Nature Reviews Chemistry}\ }\textbf {\bibinfo {volume} {3}},\ \bibinfo
  {pages} {250--260} (\bibinfo {year} {2019})}\BibitemShut {NoStop}%
\bibitem [{\citenamefont {Turner}\ \emph {et~al.}(2013)\citenamefont {Turner},
  \citenamefont {Saba}, \citenamefont {Zhang}, \citenamefont {Cumming},
  \citenamefont {Schr{\"o}der-Turk},\ and\ \citenamefont {Gu}}]{tur13}%
  \BibitemOpen
  \bibfield  {author} {\bibinfo {author} {\bibfnamefont {M.~D.}\ \bibnamefont
  {Turner}}, \bibinfo {author} {\bibfnamefont {M.}~\bibnamefont {Saba}},
  \bibinfo {author} {\bibfnamefont {Q.}~\bibnamefont {Zhang}}, \bibinfo
  {author} {\bibfnamefont {B.~P.}\ \bibnamefont {Cumming}}, \bibinfo {author}
  {\bibfnamefont {G.~E.}\ \bibnamefont {Schr{\"o}der-Turk}}, \ and\ \bibinfo
  {author} {\bibfnamefont {M.}~\bibnamefont {Gu}},\ }\bibfield  {title}
  {\enquote {\bibinfo {title} {Miniature chiral beamsplitter based on gyroid
  photonic crystals},}\ }\href@noop {} {\bibfield  {journal} {\bibinfo
  {journal} {Nature Photonics}\ }\textbf {\bibinfo {volume} {7}},\ \bibinfo
  {pages} {801--805} (\bibinfo {year} {2013})}\BibitemShut {NoStop}%
\bibitem [{\citenamefont {Chang}\ \emph {et~al.}(2018)\citenamefont {Chang},
  \citenamefont {Wieder}, \citenamefont {Schindler}, \citenamefont {Sanchez},
  \citenamefont {Belopolski}, \citenamefont {Huang}, \citenamefont {Singh},
  \citenamefont {Wu}, \citenamefont {Chang}, \citenamefont {Neupert} \emph
  {et~al.}}]{ch18}%
  \BibitemOpen
  \bibfield  {author} {\bibinfo {author} {\bibfnamefont {G.}~\bibnamefont
  {Chang}}, \bibinfo {author} {\bibfnamefont {B.~J.}\ \bibnamefont {Wieder}},
  \bibinfo {author} {\bibfnamefont {F.}~\bibnamefont {Schindler}}, \bibinfo
  {author} {\bibfnamefont {D.~S.}\ \bibnamefont {Sanchez}}, \bibinfo {author}
  {\bibfnamefont {I.}~\bibnamefont {Belopolski}}, \bibinfo {author}
  {\bibfnamefont {S.-M.}\ \bibnamefont {Huang}}, \bibinfo {author}
  {\bibfnamefont {B.}~\bibnamefont {Singh}}, \bibinfo {author} {\bibfnamefont
  {D.}~\bibnamefont {Wu}}, \bibinfo {author} {\bibfnamefont {T.-R.}\
  \bibnamefont {Chang}}, \bibinfo {author} {\bibfnamefont {T.}~\bibnamefont
  {Neupert}},  \emph {et~al.},\ }\bibfield  {title} {\enquote {\bibinfo {title}
  {Topological quantum properties of chiral crystals},}\ }\href@noop {}
  {\bibfield  {journal} {\bibinfo  {journal} {Nature materials}\ }\textbf
  {\bibinfo {volume} {17}},\ \bibinfo {pages} {978--985} (\bibinfo {year}
  {2018})}\BibitemShut {NoStop}%
\bibitem [{\citenamefont {Armitage}, \citenamefont {Mele},\ and\ \citenamefont
  {Vishwanath}(2018)}]{Armitage2018}%
  \BibitemOpen
  \bibfield  {author} {\bibinfo {author} {\bibfnamefont {N.~P.}\ \bibnamefont
  {Armitage}}, \bibinfo {author} {\bibfnamefont {E.~J.}\ \bibnamefont {Mele}},
  \ and\ \bibinfo {author} {\bibfnamefont {A.}~\bibnamefont {Vishwanath}},\
  }\bibfield  {title} {\enquote {\bibinfo {title} {{W}eyl and {D}irac
  semimetals in three-dimensional solids},}\ }\href {\doibase
  10.1103/RevModPhys.90.015001} {\bibfield  {journal} {\bibinfo  {journal}
  {Rev. Mod. Phys.}\ }\textbf {\bibinfo {volume} {90}},\ \bibinfo {pages}
  {015001} (\bibinfo {year} {2018})}\BibitemShut {NoStop}%
\bibitem [{\citenamefont {Lv}, \citenamefont {Qian},\ and\ \citenamefont
  {Ding}(2021)}]{Lv2021}%
  \BibitemOpen
  \bibfield  {author} {\bibinfo {author} {\bibfnamefont {B.~Q.}\ \bibnamefont
  {Lv}}, \bibinfo {author} {\bibfnamefont {T.}~\bibnamefont {Qian}}, \ and\
  \bibinfo {author} {\bibfnamefont {H.}~\bibnamefont {Ding}},\ }\bibfield
  {title} {\enquote {\bibinfo {title} {Experimental perspective on
  three-dimensional topological semimetals},}\ }\href {\doibase
  10.1103/RevModPhys.93.025002} {\bibfield  {journal} {\bibinfo  {journal}
  {Rev. Mod. Phys.}\ }\textbf {\bibinfo {volume} {93}},\ \bibinfo {pages}
  {025002} (\bibinfo {year} {2021})}\BibitemShut {NoStop}%
\bibitem [{\citenamefont {Chang}\ \emph {et~al.}(2017)\citenamefont {Chang},
  \citenamefont {Xu}, \citenamefont {Wieder}, \citenamefont {Sanchez},
  \citenamefont {Huang}, \citenamefont {Belopolski}, \citenamefont {Chang},
  \citenamefont {Zhang}, \citenamefont {Bansil}, \citenamefont {Lin},\ and\
  \citenamefont {Hasan}}]{Chang2017}%
  \BibitemOpen
  \bibfield  {author} {\bibinfo {author} {\bibfnamefont {G.}~\bibnamefont
  {Chang}}, \bibinfo {author} {\bibfnamefont {S.-Y.}\ \bibnamefont {Xu}},
  \bibinfo {author} {\bibfnamefont {B.~J.}\ \bibnamefont {Wieder}}, \bibinfo
  {author} {\bibfnamefont {D.~S.}\ \bibnamefont {Sanchez}}, \bibinfo {author}
  {\bibfnamefont {S.-M.}\ \bibnamefont {Huang}}, \bibinfo {author}
  {\bibfnamefont {I.}~\bibnamefont {Belopolski}}, \bibinfo {author}
  {\bibfnamefont {T.-R.}\ \bibnamefont {Chang}}, \bibinfo {author}
  {\bibfnamefont {S.}~\bibnamefont {Zhang}}, \bibinfo {author} {\bibfnamefont
  {A.}~\bibnamefont {Bansil}}, \bibinfo {author} {\bibfnamefont
  {H.}~\bibnamefont {Lin}}, \ and\ \bibinfo {author} {\bibfnamefont {M.~Z.}\
  \bibnamefont {Hasan}},\ }\bibfield  {title} {\enquote {\bibinfo {title}
  {Unconventional chiral fermions and large topological {F}ermi arcs in
  {R}h{S}i},}\ }\href {\doibase 10.1103/PhysRevLett.119.206401} {\bibfield
  {journal} {\bibinfo  {journal} {Phys. Rev. Lett.}\ }\textbf {\bibinfo
  {volume} {119}},\ \bibinfo {pages} {206401} (\bibinfo {year}
  {2017})}\BibitemShut {NoStop}%
\bibitem [{\citenamefont {Bradlyn}\ \emph {et~al.}(2016)\citenamefont
  {Bradlyn}, \citenamefont {Cano}, \citenamefont {Wang}, \citenamefont
  {Vergniory}, \citenamefont {Felser}, \citenamefont {Cava},\ and\
  \citenamefont {Bernevig}}]{Bradlyn2016}%
  \BibitemOpen
  \bibfield  {author} {\bibinfo {author} {\bibfnamefont {B.}~\bibnamefont
  {Bradlyn}}, \bibinfo {author} {\bibfnamefont {J.}~\bibnamefont {Cano}},
  \bibinfo {author} {\bibfnamefont {Z.}~\bibnamefont {Wang}}, \bibinfo {author}
  {\bibfnamefont {M.~G.}\ \bibnamefont {Vergniory}}, \bibinfo {author}
  {\bibfnamefont {C.}~\bibnamefont {Felser}}, \bibinfo {author} {\bibfnamefont
  {R.~J.}\ \bibnamefont {Cava}}, \ and\ \bibinfo {author} {\bibfnamefont
  {B.~A.}\ \bibnamefont {Bernevig}},\ }\bibfield  {title} {\enquote {\bibinfo
  {title} {Beyond {D}irac and {W}eyl fermions: Unconventional quasiparticles in
  conventional crystals},}\ }\href {\doibase 10.1126/science.aaf5037}
  {\bibfield  {journal} {\bibinfo  {journal} {Science}\ }\textbf {\bibinfo
  {volume} {353}},\ \bibinfo {pages} {6299} (\bibinfo {year}
  {2016})}\BibitemShut {NoStop}%
\bibitem [{\citenamefont {Takane}\ \emph {et~al.}(2019)\citenamefont {Takane},
  \citenamefont {Wang}, \citenamefont {Souma}, \citenamefont {Nakayama},
  \citenamefont {Nakamura}, \citenamefont {Oinuma}, \citenamefont {Nakata},
  \citenamefont {Iwasawa}, \citenamefont {Cacho}, \citenamefont {Kim},
  \citenamefont {Horiba}, \citenamefont {Kumigashira}, \citenamefont
  {Takahashi}, \citenamefont {Ando},\ and\ \citenamefont {Sato}}]{Takane2019}%
  \BibitemOpen
  \bibfield  {author} {\bibinfo {author} {\bibfnamefont {D.}~\bibnamefont
  {Takane}}, \bibinfo {author} {\bibfnamefont {Z.}~\bibnamefont {Wang}},
  \bibinfo {author} {\bibfnamefont {S.}~\bibnamefont {Souma}}, \bibinfo
  {author} {\bibfnamefont {K.}~\bibnamefont {Nakayama}}, \bibinfo {author}
  {\bibfnamefont {T.}~\bibnamefont {Nakamura}}, \bibinfo {author}
  {\bibfnamefont {H.}~\bibnamefont {Oinuma}}, \bibinfo {author} {\bibfnamefont
  {Y.}~\bibnamefont {Nakata}}, \bibinfo {author} {\bibfnamefont
  {H.}~\bibnamefont {Iwasawa}}, \bibinfo {author} {\bibfnamefont
  {C.}~\bibnamefont {Cacho}}, \bibinfo {author} {\bibfnamefont
  {T.}~\bibnamefont {Kim}}, \bibinfo {author} {\bibfnamefont {K.}~\bibnamefont
  {Horiba}}, \bibinfo {author} {\bibfnamefont {H.}~\bibnamefont {Kumigashira}},
  \bibinfo {author} {\bibfnamefont {T.}~\bibnamefont {Takahashi}}, \bibinfo
  {author} {\bibfnamefont {Y.}~\bibnamefont {Ando}}, \ and\ \bibinfo {author}
  {\bibfnamefont {T.}~\bibnamefont {Sato}},\ }\bibfield  {title} {\enquote
  {\bibinfo {title} {Observation of chiral fermions with a large topological
  charge and associated {F}ermi-arc surface states in {C}o{S}i},}\ }\href
  {\doibase 10.1103/PhysRevLett.122.076402} {\bibfield  {journal} {\bibinfo
  {journal} {Phys. Rev. Lett.}\ }\textbf {\bibinfo {volume} {122}},\ \bibinfo
  {pages} {076402} (\bibinfo {year} {2019})}\BibitemShut {NoStop}%
\bibitem [{\citenamefont {Rao}\ \emph {et~al.}(2019)\citenamefont {Rao},
  \citenamefont {Li}, \citenamefont {Zhang}, \citenamefont {Tian},
  \citenamefont {Chenghe}, \citenamefont {Fu}, \citenamefont {Tang},
  \citenamefont {Wang}, \citenamefont {Li}, \citenamefont {Fan}, \citenamefont
  {Li}, \citenamefont {Huang}, \citenamefont {Lei}, \citenamefont {Long},
  \citenamefont {Fang}, \citenamefont {Weng}, \citenamefont {Shi},
  \citenamefont {Sun}, \citenamefont {Qian},\ and\ \citenamefont
  {Ding}}]{Rao2019}%
  \BibitemOpen
  \bibfield  {author} {\bibinfo {author} {\bibfnamefont {Z.}~\bibnamefont
  {Rao}}, \bibinfo {author} {\bibfnamefont {H.}~\bibnamefont {Li}}, \bibinfo
  {author} {\bibfnamefont {T.}~\bibnamefont {Zhang}}, \bibinfo {author}
  {\bibfnamefont {S.-J.}\ \bibnamefont {Tian}}, \bibinfo {author}
  {\bibfnamefont {L.}~\bibnamefont {Chenghe}}, \bibinfo {author} {\bibfnamefont
  {B.}~\bibnamefont {Fu}}, \bibinfo {author} {\bibfnamefont {C.}~\bibnamefont
  {Tang}}, \bibinfo {author} {\bibfnamefont {L.}~\bibnamefont {Wang}}, \bibinfo
  {author} {\bibfnamefont {Z.}~\bibnamefont {Li}}, \bibinfo {author}
  {\bibfnamefont {W.}~\bibnamefont {Fan}}, \bibinfo {author} {\bibfnamefont
  {J.}~\bibnamefont {Li}}, \bibinfo {author} {\bibfnamefont {Y.}~\bibnamefont
  {Huang}}, \bibinfo {author} {\bibfnamefont {H.}~\bibnamefont {Lei}}, \bibinfo
  {author} {\bibfnamefont {Y.}~\bibnamefont {Long}}, \bibinfo {author}
  {\bibfnamefont {C.}~\bibnamefont {Fang}}, \bibinfo {author} {\bibfnamefont
  {H.}~\bibnamefont {Weng}}, \bibinfo {author} {\bibfnamefont {Y.}~\bibnamefont
  {Shi}}, \bibinfo {author} {\bibfnamefont {Y.}~\bibnamefont {Sun}}, \bibinfo
  {author} {\bibfnamefont {T.}~\bibnamefont {Qian}}, \ and\ \bibinfo {author}
  {\bibfnamefont {H.}~\bibnamefont {Ding}},\ }\bibfield  {title} {\enquote
  {\bibinfo {title} {Observation of unconventional chiral fermions with long
  {F}ermi arcs in {C}o{S}i},}\ }\href {\doibase 10.1038/s41586-019-1031-8}
  {\bibfield  {journal} {\bibinfo  {journal} {Nature}\ }\textbf {\bibinfo
  {volume} {567}},\ \bibinfo {pages} {496} (\bibinfo {year}
  {2019})}\BibitemShut {NoStop}%
\bibitem [{\citenamefont {Sanchez}\ \emph {et~al.}(2019)\citenamefont
  {Sanchez}, \citenamefont {Belopolski}, \citenamefont {Cochran}, \citenamefont
  {Xu}, \citenamefont {Yin}, \citenamefont {Chang}, \citenamefont {Xie},
  \citenamefont {Manna}, \citenamefont {S\"{u}{\ss}}, \citenamefont {Huang},
  \citenamefont {Alidoust}, \citenamefont {Multer}, \citenamefont {Zhang},
  \citenamefont {Shumiya}, \citenamefont {Wang}, \citenamefont {Wang},
  \citenamefont {Chang}, \citenamefont {Felser}, \citenamefont {Xu},\ and\
  \citenamefont {Hasan}}]{Sanchez2019}%
  \BibitemOpen
  \bibfield  {author} {\bibinfo {author} {\bibfnamefont {D.~S.}\ \bibnamefont
  {Sanchez}}, \bibinfo {author} {\bibfnamefont {I.}~\bibnamefont {Belopolski}},
  \bibinfo {author} {\bibfnamefont {T.~A.}\ \bibnamefont {Cochran}}, \bibinfo
  {author} {\bibfnamefont {X.}~\bibnamefont {Xu}}, \bibinfo {author}
  {\bibfnamefont {J.-X.}\ \bibnamefont {Yin}}, \bibinfo {author} {\bibfnamefont
  {G.}~\bibnamefont {Chang}}, \bibinfo {author} {\bibfnamefont
  {W.}~\bibnamefont {Xie}}, \bibinfo {author} {\bibfnamefont {K.}~\bibnamefont
  {Manna}}, \bibinfo {author} {\bibfnamefont {V.}~\bibnamefont {S\"{u}{\ss}}},
  \bibinfo {author} {\bibfnamefont {C.-Y.}\ \bibnamefont {Huang}}, \bibinfo
  {author} {\bibfnamefont {N.}~\bibnamefont {Alidoust}}, \bibinfo {author}
  {\bibfnamefont {D.}~\bibnamefont {Multer}}, \bibinfo {author} {\bibfnamefont
  {S.}~\bibnamefont {Zhang}}, \bibinfo {author} {\bibfnamefont
  {N.}~\bibnamefont {Shumiya}}, \bibinfo {author} {\bibfnamefont
  {X.}~\bibnamefont {Wang}}, \bibinfo {author} {\bibfnamefont {G.-Q.}\
  \bibnamefont {Wang}}, \bibinfo {author} {\bibfnamefont {T.-R.}\ \bibnamefont
  {Chang}}, \bibinfo {author} {\bibfnamefont {C.}~\bibnamefont {Felser}},
  \bibinfo {author} {\bibfnamefont {S.-Y.}\ \bibnamefont {Xu}}, \ and\ \bibinfo
  {author} {\bibfnamefont {M.~Z.}\ \bibnamefont {Hasan}},\ }\bibfield  {title}
  {\enquote {\bibinfo {title} {Topological chiral crystals with helicoid-arc
  quantum states},}\ }\href {\doibase 10.1038/s41586-019-1037-2} {\bibfield
  {journal} {\bibinfo  {journal} {Nature}\ }\textbf {\bibinfo {volume} {567}},\
  \bibinfo {pages} {500} (\bibinfo {year} {2019})}\BibitemShut {NoStop}%
\bibitem [{\citenamefont {Min}\ \emph {et~al.}(2019)\citenamefont {Min},
  \citenamefont {Bentmann}, \citenamefont {Neu}, \citenamefont {Eck},
  \citenamefont {Moser}, \citenamefont {Figgemeier}, \citenamefont
  {\"{U}nzelmann}, \citenamefont {Kissner}, \citenamefont {Lutz}, \citenamefont
  {Koch}, \citenamefont {Jozwiak}, \citenamefont {Bostwick}, \citenamefont
  {Rotenberg}, \citenamefont {Thomale}, \citenamefont {Sangiovanni},
  \citenamefont {Siegrist}, \citenamefont {Di~Sante},\ and\ \citenamefont
  {Reinert}}]{Min2019}%
  \BibitemOpen
  \bibfield  {author} {\bibinfo {author} {\bibfnamefont {C.-H.}\ \bibnamefont
  {Min}}, \bibinfo {author} {\bibfnamefont {H.}~\bibnamefont {Bentmann}},
  \bibinfo {author} {\bibfnamefont {J.~N.}\ \bibnamefont {Neu}}, \bibinfo
  {author} {\bibfnamefont {P.}~\bibnamefont {Eck}}, \bibinfo {author}
  {\bibfnamefont {S.}~\bibnamefont {Moser}}, \bibinfo {author} {\bibfnamefont
  {T.}~\bibnamefont {Figgemeier}}, \bibinfo {author} {\bibfnamefont
  {M.}~\bibnamefont {\"{U}nzelmann}}, \bibinfo {author} {\bibfnamefont
  {K.}~\bibnamefont {Kissner}}, \bibinfo {author} {\bibfnamefont
  {P.}~\bibnamefont {Lutz}}, \bibinfo {author} {\bibfnamefont {R.~J.}\
  \bibnamefont {Koch}}, \bibinfo {author} {\bibfnamefont {C.}~\bibnamefont
  {Jozwiak}}, \bibinfo {author} {\bibfnamefont {A.}~\bibnamefont {Bostwick}},
  \bibinfo {author} {\bibfnamefont {E.}~\bibnamefont {Rotenberg}}, \bibinfo
  {author} {\bibfnamefont {R.}~\bibnamefont {Thomale}}, \bibinfo {author}
  {\bibfnamefont {G.}~\bibnamefont {Sangiovanni}}, \bibinfo {author}
  {\bibfnamefont {T.}~\bibnamefont {Siegrist}}, \bibinfo {author}
  {\bibfnamefont {D.}~\bibnamefont {Di~Sante}}, \ and\ \bibinfo {author}
  {\bibfnamefont {F.}~\bibnamefont {Reinert}},\ }\bibfield  {title} {\enquote
  {\bibinfo {title} {Orbital fingerprint of topological {F}ermi arcs in the
  {W}eyl semimetal {T}a{P}},}\ }\href {\doibase 10.1103/PhysRevLett.122.116402}
  {\bibfield  {journal} {\bibinfo  {journal} {Phys. Rev. Lett.}\ }\textbf
  {\bibinfo {volume} {122}},\ \bibinfo {pages} {116402} (\bibinfo {year}
  {2019})}\BibitemShut {NoStop}%
\bibitem [{\citenamefont {\"{U}nzelmann}\ \emph {et~al.}(2021)\citenamefont
  {\"{U}nzelmann}, \citenamefont {Bentmann}, \citenamefont {Figgemeier},
  \citenamefont {Eck}, \citenamefont {Neu}, \citenamefont {Geldiyev},
  \citenamefont {Diekmann}, \citenamefont {Rohlf}, \citenamefont {Buck},
  \citenamefont {Hoesch}, \citenamefont {Kalläne}, \citenamefont {Rossnagel},
  \citenamefont {Thomale}, \citenamefont {Siegrist}, \citenamefont
  {Sangiovanni}, \citenamefont {Sante},\ and\ \citenamefont
  {Reinert}}]{Uenzelmann2021}%
  \BibitemOpen
  \bibfield  {author} {\bibinfo {author} {\bibfnamefont {M.}~\bibnamefont
  {\"{U}nzelmann}}, \bibinfo {author} {\bibfnamefont {H.}~\bibnamefont
  {Bentmann}}, \bibinfo {author} {\bibfnamefont {T.}~\bibnamefont
  {Figgemeier}}, \bibinfo {author} {\bibfnamefont {P.}~\bibnamefont {Eck}},
  \bibinfo {author} {\bibfnamefont {J.~N.}\ \bibnamefont {Neu}}, \bibinfo
  {author} {\bibfnamefont {B.}~\bibnamefont {Geldiyev}}, \bibinfo {author}
  {\bibfnamefont {F.}~\bibnamefont {Diekmann}}, \bibinfo {author}
  {\bibfnamefont {S.}~\bibnamefont {Rohlf}}, \bibinfo {author} {\bibfnamefont
  {J.}~\bibnamefont {Buck}}, \bibinfo {author} {\bibfnamefont {M.}~\bibnamefont
  {Hoesch}}, \bibinfo {author} {\bibfnamefont {M.}~\bibnamefont {Kalläne}},
  \bibinfo {author} {\bibfnamefont {K.}~\bibnamefont {Rossnagel}}, \bibinfo
  {author} {\bibfnamefont {R.}~\bibnamefont {Thomale}}, \bibinfo {author}
  {\bibfnamefont {T.}~\bibnamefont {Siegrist}}, \bibinfo {author}
  {\bibfnamefont {G.}~\bibnamefont {Sangiovanni}}, \bibinfo {author}
  {\bibfnamefont {D.~D.}\ \bibnamefont {Sante}}, \ and\ \bibinfo {author}
  {\bibfnamefont {F.}~\bibnamefont {Reinert}},\ }\bibfield  {title} {\enquote
  {\bibinfo {title} {Momentum-space signatures of {B}erry flux monopoles in the
  {W}eyl semimetal {T}a{A}s},}\ }\href {\doibase 10.1038/s41467-021-23727-3}
  {\bibfield  {journal} {\bibinfo  {journal} {Nat. Commun.}\ }\textbf {\bibinfo
  {volume} {12}},\ \bibinfo {pages} {3650} (\bibinfo {year}
  {2021})}\BibitemShut {NoStop}%
\bibitem [{\citenamefont {Park}\ \emph
  {et~al.}(2012{\natexlab{a}})\citenamefont {Park}, \citenamefont {Han},
  \citenamefont {Kim}, \citenamefont {Koh}, \citenamefont {Kim}, \citenamefont
  {Lee}, \citenamefont {Choi}, \citenamefont {Han}, \citenamefont {Lee},
  \citenamefont {Hur}, \citenamefont {Arita}, \citenamefont {Shimada},
  \citenamefont {Namatame},\ and\ \citenamefont {Taniguchi}}]{Park2012}%
  \BibitemOpen
  \bibfield  {author} {\bibinfo {author} {\bibfnamefont {S.~R.}\ \bibnamefont
  {Park}}, \bibinfo {author} {\bibfnamefont {J.}~\bibnamefont {Han}}, \bibinfo
  {author} {\bibfnamefont {C.}~\bibnamefont {Kim}}, \bibinfo {author}
  {\bibfnamefont {Y.~Y.}\ \bibnamefont {Koh}}, \bibinfo {author} {\bibfnamefont
  {C.}~\bibnamefont {Kim}}, \bibinfo {author} {\bibfnamefont {H.}~\bibnamefont
  {Lee}}, \bibinfo {author} {\bibfnamefont {H.~J.}\ \bibnamefont {Choi}},
  \bibinfo {author} {\bibfnamefont {J.~H.}\ \bibnamefont {Han}}, \bibinfo
  {author} {\bibfnamefont {K.~D.}\ \bibnamefont {Lee}}, \bibinfo {author}
  {\bibfnamefont {N.~J.}\ \bibnamefont {Hur}}, \bibinfo {author} {\bibfnamefont
  {M.}~\bibnamefont {Arita}}, \bibinfo {author} {\bibfnamefont
  {K.}~\bibnamefont {Shimada}}, \bibinfo {author} {\bibfnamefont
  {H.}~\bibnamefont {Namatame}}, \ and\ \bibinfo {author} {\bibfnamefont
  {M.}~\bibnamefont {Taniguchi}},\ }\bibfield  {title} {\enquote {\bibinfo
  {title} {Chiral orbital-angular momentum in the surface states of
  {B}i$_{2}${S}e$_{3}$},}\ }\href {\doibase 10.1103/PhysRevLett.108.046805}
  {\bibfield  {journal} {\bibinfo  {journal} {Phys. Rev. Lett.}\ }\textbf
  {\bibinfo {volume} {108}},\ \bibinfo {pages} {046805} (\bibinfo {year}
  {2012}{\natexlab{a}})}\BibitemShut {NoStop}%
\bibitem [{\citenamefont {Park}\ \emph
  {et~al.}(2012{\natexlab{b}})\citenamefont {Park}, \citenamefont {Kim},
  \citenamefont {Rhim},\ and\ \citenamefont {Han}}]{Park2012a}%
  \BibitemOpen
  \bibfield  {author} {\bibinfo {author} {\bibfnamefont {J.-H.}\ \bibnamefont
  {Park}}, \bibinfo {author} {\bibfnamefont {C.~H.}\ \bibnamefont {Kim}},
  \bibinfo {author} {\bibfnamefont {J.-W.}\ \bibnamefont {Rhim}}, \ and\
  \bibinfo {author} {\bibfnamefont {J.~H.}\ \bibnamefont {Han}},\ }\bibfield
  {title} {\enquote {\bibinfo {title} {Orbital {R}ashba effect and its
  detection by circular dichroism angle-resolved photoemission spectroscopy},}\
  }\href {\doibase 10.1103/PhysRevB.85.195401} {\bibfield  {journal} {\bibinfo
  {journal} {Phys. Rev. B}\ }\textbf {\bibinfo {volume} {85}},\ \bibinfo
  {pages} {195401} (\bibinfo {year} {2012}{\natexlab{b}})}\BibitemShut
  {NoStop}%
\bibitem [{\citenamefont {Sch\"{u}ler}\ \emph {et~al.}(2020)\citenamefont
  {Sch\"{u}ler}, \citenamefont {Giovannini}, \citenamefont {Hübener},
  \citenamefont {Rubio}, \citenamefont {Sentef},\ and\ \citenamefont
  {Werner}}]{Schueler2020}%
  \BibitemOpen
  \bibfield  {author} {\bibinfo {author} {\bibfnamefont {M.}~\bibnamefont
  {Sch\"{u}ler}}, \bibinfo {author} {\bibfnamefont {U.~D.}\ \bibnamefont
  {Giovannini}}, \bibinfo {author} {\bibfnamefont {H.}~\bibnamefont
  {Hübener}}, \bibinfo {author} {\bibfnamefont {A.}~\bibnamefont {Rubio}},
  \bibinfo {author} {\bibfnamefont {M.~A.}\ \bibnamefont {Sentef}}, \ and\
  \bibinfo {author} {\bibfnamefont {P.}~\bibnamefont {Werner}},\ }\bibfield
  {title} {\enquote {\bibinfo {title} {Local {B}erry curvature signatures in
  dichroic angle-resolved photoelectron spectroscopy from two-dimensional
  materials},}\ }\href@noop {} {\bibfield  {journal} {\bibinfo  {journal} {Sci.
  Adv.}\ }\textbf {\bibinfo {volume} {6}},\ \bibinfo {pages} {eaay2730}
  (\bibinfo {year} {2020})}\BibitemShut {NoStop}%
\bibitem [{\citenamefont {Yang}\ \emph {et~al.}(2023)\citenamefont {Yang},
  \citenamefont {Xiao}, \citenamefont {Robredo}, \citenamefont {Vergniory},
  \citenamefont {Yan},\ and\ \citenamefont {Felser}}]{Yang2023}%
  \BibitemOpen
  \bibfield  {author} {\bibinfo {author} {\bibfnamefont {Q.}~\bibnamefont
  {Yang}}, \bibinfo {author} {\bibfnamefont {J.}~\bibnamefont {Xiao}}, \bibinfo
  {author} {\bibfnamefont {I.}~\bibnamefont {Robredo}}, \bibinfo {author}
  {\bibfnamefont {M.~G.}\ \bibnamefont {Vergniory}}, \bibinfo {author}
  {\bibfnamefont {B.}~\bibnamefont {Yan}}, \ and\ \bibinfo {author}
  {\bibfnamefont {C.}~\bibnamefont {Felser}},\ }\bibfield  {title} {\enquote
  {\bibinfo {title} {Monopole-like orbital-momentum locking and the induced
  orbital transport in topological chiral semimetals},}\ }\href {\doibase
  10.1073/pnas.2305541120} {\bibfield  {journal} {\bibinfo  {journal}
  {Proceedings of the National Academy of Sciences}\ }\textbf {\bibinfo
  {volume} {120}},\ \bibinfo {pages} {e2305541120} (\bibinfo {year}
  {2023})}\BibitemShut {NoStop}%
\bibitem [{\citenamefont {Wiemann}\ \emph {et~al.}(2011)\citenamefont
  {Wiemann}, \citenamefont {Patt}, \citenamefont {Krug}, \citenamefont {Weber},
  \citenamefont {Escher}, \citenamefont {Merkel},\ and\ \citenamefont
  {Schneider}}]{Wiemann2011}%
  \BibitemOpen
  \bibfield  {author} {\bibinfo {author} {\bibfnamefont {C.}~\bibnamefont
  {Wiemann}}, \bibinfo {author} {\bibfnamefont {M.}~\bibnamefont {Patt}},
  \bibinfo {author} {\bibfnamefont {I.~P.}\ \bibnamefont {Krug}}, \bibinfo
  {author} {\bibfnamefont {N.~B.}\ \bibnamefont {Weber}}, \bibinfo {author}
  {\bibfnamefont {M.}~\bibnamefont {Escher}}, \bibinfo {author} {\bibfnamefont
  {M.}~\bibnamefont {Merkel}}, \ and\ \bibinfo {author} {\bibfnamefont {C.~M.}\
  \bibnamefont {Schneider}},\ }\bibfield  {title} {\enquote {\bibinfo {title}
  {A new nanospectroscopy tool with synchrotron radiation: Nanoesca@
  {E}lettra},}\ }\href {\doibase 10.1380/ejssnt.2011.395} {\bibfield  {journal}
  {\bibinfo  {journal} {e-J. Surf. Sci. Nanotechnol.}\ }\textbf {\bibinfo
  {volume} {9}},\ \bibinfo {pages} {395--399} (\bibinfo {year}
  {2011})}\BibitemShut {NoStop}%
\bibitem [{\citenamefont {Tusche}, \citenamefont {Krasyuk},\ and\ \citenamefont
  {Kirschner}(2015)}]{tus15}%
  \BibitemOpen
  \bibfield  {author} {\bibinfo {author} {\bibfnamefont {C.}~\bibnamefont
  {Tusche}}, \bibinfo {author} {\bibfnamefont {A.}~\bibnamefont {Krasyuk}}, \
  and\ \bibinfo {author} {\bibfnamefont {J.}~\bibnamefont {Kirschner}},\
  }\bibfield  {title} {\enquote {\bibinfo {title} {Spin resolved bandstructure
  imaging with a high resolution momentum microscope},}\ }\href {\doibase
  10.1016/j.ultramic.2015.03.020} {\bibfield  {journal} {\bibinfo  {journal}
  {Ultramicroscopy}\ }\textbf {\bibinfo {volume} {159}},\ \bibinfo {pages}
  {520--529} (\bibinfo {year} {2015})}\BibitemShut {NoStop}%
\bibitem [{\citenamefont {Tusche}\ \emph
  {et~al.}(2019{\natexlab{a}})\citenamefont {Tusche}, \citenamefont {Chen},
  \citenamefont {Schneider},\ and\ \citenamefont {Kirschner}}]{Tusche2019}%
  \BibitemOpen
  \bibfield  {author} {\bibinfo {author} {\bibfnamefont {C.}~\bibnamefont
  {Tusche}}, \bibinfo {author} {\bibfnamefont {Y.-J.}\ \bibnamefont {Chen}},
  \bibinfo {author} {\bibfnamefont {C.~M.}\ \bibnamefont {Schneider}}, \ and\
  \bibinfo {author} {\bibfnamefont {J.}~\bibnamefont {Kirschner}},\ }\bibfield
  {title} {\enquote {\bibinfo {title} {Imaging properties of hemispherical
  electrostatic energy analyzers for high resolution momentum microscopy},}\
  }\href {\doibase 10.1016/j.ultramic.2019.112815} {\bibfield  {journal}
  {\bibinfo  {journal} {Ultramicroscopy}\ }\textbf {\bibinfo {volume} {206}},\
  \bibinfo {pages} {112815} (\bibinfo {year} {2019}{\natexlab{a}})}\BibitemShut
  {NoStop}%
\bibitem [{\citenamefont {Liu}\ \emph {et~al.}(2011)\citenamefont {Liu},
  \citenamefont {Bian}, \citenamefont {Miller},\ and\ \citenamefont
  {Chiang}}]{Liu2011a}%
  \BibitemOpen
  \bibfield  {author} {\bibinfo {author} {\bibfnamefont {Y.}~\bibnamefont
  {Liu}}, \bibinfo {author} {\bibfnamefont {G.}~\bibnamefont {Bian}}, \bibinfo
  {author} {\bibfnamefont {T.}~\bibnamefont {Miller}}, \ and\ \bibinfo {author}
  {\bibfnamefont {T.-C.}\ \bibnamefont {Chiang}},\ }\bibfield  {title}
  {\enquote {\bibinfo {title} {Visualizing electronic chirality and {B}erry
  phases in graphene systems using photoemission with circularly polarized
  light},}\ }\href {\doibase 10.1103/PhysRevLett.107.166803} {\bibfield
  {journal} {\bibinfo  {journal} {Phys. Rev. Lett.}\ }\textbf {\bibinfo
  {volume} {107}},\ \bibinfo {pages} {166803} (\bibinfo {year}
  {2011})}\BibitemShut {NoStop}%
\bibitem [{\citenamefont {Hagiwara}\ \emph {et~al.}(2022)\citenamefont
  {Hagiwara}, \citenamefont {R{\"u}{\ss}mann}, \citenamefont {Tan},
  \citenamefont {Chen}, \citenamefont {Ueno}, \citenamefont {Feyer},
  \citenamefont {Zamborlini}, \citenamefont {Jugovac}, \citenamefont {Suga},
  \citenamefont {Bl{\"u}gel}, \citenamefont {Schneider},\ and\ \citenamefont
  {Tusche}}]{Hagiwara2022}%
  \BibitemOpen
  \bibfield  {author} {\bibinfo {author} {\bibfnamefont {K.}~\bibnamefont
  {Hagiwara}}, \bibinfo {author} {\bibfnamefont {P.}~\bibnamefont
  {R{\"u}{\ss}mann}}, \bibinfo {author} {\bibfnamefont {X.~L.}\ \bibnamefont
  {Tan}}, \bibinfo {author} {\bibfnamefont {Y.-J.}\ \bibnamefont {Chen}},
  \bibinfo {author} {\bibfnamefont {K.}~\bibnamefont {Ueno}}, \bibinfo {author}
  {\bibfnamefont {V.}~\bibnamefont {Feyer}}, \bibinfo {author} {\bibfnamefont
  {G.}~\bibnamefont {Zamborlini}}, \bibinfo {author} {\bibfnamefont
  {M.}~\bibnamefont {Jugovac}}, \bibinfo {author} {\bibfnamefont
  {S.}~\bibnamefont {Suga}}, \bibinfo {author} {\bibfnamefont {S.}~\bibnamefont
  {Bl{\"u}gel}}, \bibinfo {author} {\bibfnamefont {C.~M.}\ \bibnamefont
  {Schneider}}, \ and\ \bibinfo {author} {\bibfnamefont {C.}~\bibnamefont
  {Tusche}},\ }\bibfield  {title} {\enquote {\bibinfo {title} {Link between
  {W}eyl-fermion chirality and spin texture},}\ }\href@noop {} {\bibfield
  {journal} {\bibinfo  {journal} {arXiv preprint arXiv:2205.15252}\ } (\bibinfo
  {year} {2022})}\BibitemShut {NoStop}%
\bibitem [{\citenamefont {Fecher}, \citenamefont {K{\"u}bler},\ and\
  \citenamefont {Felser}(2022)}]{fe22}%
  \BibitemOpen
  \bibfield  {author} {\bibinfo {author} {\bibfnamefont {G.~H.}\ \bibnamefont
  {Fecher}}, \bibinfo {author} {\bibfnamefont {J.}~\bibnamefont {K{\"u}bler}},
  \ and\ \bibinfo {author} {\bibfnamefont {C.}~\bibnamefont {Felser}},\
  }\bibfield  {title} {\enquote {\bibinfo {title} {Chirality in the solid
  state: Chiral crystal structures in chiral and achiral space groups},}\
  }\href@noop {} {\bibfield  {journal} {\bibinfo  {journal} {Materials}\
  }\textbf {\bibinfo {volume} {15}},\ \bibinfo {pages} {5812} (\bibinfo {year}
  {2022})}\BibitemShut {NoStop}%
\bibitem [{\citenamefont {Sch{\"u}ler}\ and\ \citenamefont
  {Beaulieu}(2022)}]{sc22}%
  \BibitemOpen
  \bibfield  {author} {\bibinfo {author} {\bibfnamefont {M.}~\bibnamefont
  {Sch{\"u}ler}}\ and\ \bibinfo {author} {\bibfnamefont {S.}~\bibnamefont
  {Beaulieu}},\ }\bibfield  {title} {\enquote {\bibinfo {title} {Probing
  topological floquet states in {W}{S}e$_2$ using circular dichroism in
  time-and angle-resolved photoemission spectroscopy},}\ }\href@noop {}
  {\bibfield  {journal} {\bibinfo  {journal} {Communications Physics}\ }\textbf
  {\bibinfo {volume} {5}},\ \bibinfo {pages} {164} (\bibinfo {year}
  {2022})}\BibitemShut {NoStop}%
\bibitem [{\citenamefont {Go}\ \emph {et~al.}(2021{\natexlab{a}})\citenamefont
  {Go}, \citenamefont {Jo}, \citenamefont {Lee}, \citenamefont {Klaeui},\ and\
  \citenamefont {Mokrousov}}]{Go2021}%
  \BibitemOpen
  \bibfield  {author} {\bibinfo {author} {\bibfnamefont {D.}~\bibnamefont
  {Go}}, \bibinfo {author} {\bibfnamefont {D.}~\bibnamefont {Jo}}, \bibinfo
  {author} {\bibfnamefont {H.-W.}\ \bibnamefont {Lee}}, \bibinfo {author}
  {\bibfnamefont {M.}~\bibnamefont {Klaeui}}, \ and\ \bibinfo {author}
  {\bibfnamefont {Y.}~\bibnamefont {Mokrousov}},\ }\bibfield  {title} {\enquote
  {\bibinfo {title} {Orbitronics: Orbital currents in solids},}\ }\href
  {\doibase 10.1209/0295-5075/ac2653} {\bibfield  {journal} {\bibinfo
  {journal} {Europhys. Lett}\ }\textbf {\bibinfo {volume} {135}},\ \bibinfo
  {pages} {37001} (\bibinfo {year} {2021}{\natexlab{a}})}\BibitemShut {NoStop}%
\bibitem [{\citenamefont {Sch{\"o}nhense}(1990)}]{Schoenhense1990}%
  \BibitemOpen
  \bibfield  {author} {\bibinfo {author} {\bibfnamefont {G.}~\bibnamefont
  {Sch{\"o}nhense}},\ }\bibfield  {title} {\enquote {\bibinfo {title} {Circular
  dichroism and spin polarization in photoemission from adsorbates and
  non-magnetic solids},}\ }\href {\doibase 10.1088/0031-8949/1990/t31/035}
  {\bibfield  {journal} {\bibinfo  {journal} {Phys. Scr.}\ }\textbf {\bibinfo
  {volume} {1990}},\ \bibinfo {pages} {255} (\bibinfo {year}
  {1990})}\BibitemShut {NoStop}%
\bibitem [{\citenamefont {Tan}\ \emph {et~al.}(2023)\citenamefont {Tan},
  \citenamefont {Hagiwara}, \citenamefont {Chen}, \citenamefont {Schusser},
  \citenamefont {Cojocariu}, \citenamefont {Baranowski}, \citenamefont {Feyer},
  \citenamefont {Min{\'a}r}, \citenamefont {Schneider},\ and\ \citenamefont
  {Tusche}}]{ta23}%
  \BibitemOpen
  \bibfield  {author} {\bibinfo {author} {\bibfnamefont {X.~L.}\ \bibnamefont
  {Tan}}, \bibinfo {author} {\bibfnamefont {K.}~\bibnamefont {Hagiwara}},
  \bibinfo {author} {\bibfnamefont {Y.-J.}\ \bibnamefont {Chen}}, \bibinfo
  {author} {\bibfnamefont {J.}~\bibnamefont {Schusser}}, \bibinfo {author}
  {\bibfnamefont {I.}~\bibnamefont {Cojocariu}}, \bibinfo {author}
  {\bibfnamefont {D.}~\bibnamefont {Baranowski}}, \bibinfo {author}
  {\bibfnamefont {V.}~\bibnamefont {Feyer}}, \bibinfo {author} {\bibfnamefont
  {J.}~\bibnamefont {Min{\'a}r}}, \bibinfo {author} {\bibfnamefont {C.~M.}\
  \bibnamefont {Schneider}}, \ and\ \bibinfo {author} {\bibfnamefont
  {C.}~\bibnamefont {Tusche}},\ }\bibfield  {title} {\enquote {\bibinfo {title}
  {Soft x-ray {F}ermi surface tomography of {P}alladium and {R}hodium via
  momentum microscopy},}\ }\href@noop {} {\bibfield  {journal} {\bibinfo
  {journal} {Ultramicroscopy}\ }\textbf {\bibinfo {volume} {253}},\ \bibinfo
  {pages} {113820} (\bibinfo {year} {2023})}\BibitemShut {NoStop}%
\bibitem [{\citenamefont {Tusche}\ \emph {et~al.}(2018)\citenamefont {Tusche},
  \citenamefont {Ellguth}, \citenamefont {Feyer}, \citenamefont {Krasyuk},
  \citenamefont {Wiemann}, \citenamefont {Henk}, \citenamefont {Schneider},\
  and\ \citenamefont {Kirschner}}]{tu18}%
  \BibitemOpen
  \bibfield  {author} {\bibinfo {author} {\bibfnamefont {C.}~\bibnamefont
  {Tusche}}, \bibinfo {author} {\bibfnamefont {M.}~\bibnamefont {Ellguth}},
  \bibinfo {author} {\bibfnamefont {V.}~\bibnamefont {Feyer}}, \bibinfo
  {author} {\bibfnamefont {A.}~\bibnamefont {Krasyuk}}, \bibinfo {author}
  {\bibfnamefont {C.}~\bibnamefont {Wiemann}}, \bibinfo {author} {\bibfnamefont
  {J.}~\bibnamefont {Henk}}, \bibinfo {author} {\bibfnamefont {C.~M.}\
  \bibnamefont {Schneider}}, \ and\ \bibinfo {author} {\bibfnamefont
  {J.}~\bibnamefont {Kirschner}},\ }\bibfield  {title} {\enquote {\bibinfo
  {title} {Nonlocal electron correlations in an itinerant ferromagnet},}\
  }\href@noop {} {\bibfield  {journal} {\bibinfo  {journal} {Nature
  Communications}\ }\textbf {\bibinfo {volume} {9}},\ \bibinfo {pages} {3727}
  (\bibinfo {year} {2018})}\BibitemShut {NoStop}%
\bibitem [{\citenamefont {Park}\ \emph {et~al.}(2011)\citenamefont {Park},
  \citenamefont {Kim}, \citenamefont {Yu}, \citenamefont {Han},\ and\
  \citenamefont {Kim}}]{Park2011}%
  \BibitemOpen
  \bibfield  {author} {\bibinfo {author} {\bibfnamefont {S.~R.}\ \bibnamefont
  {Park}}, \bibinfo {author} {\bibfnamefont {C.~H.}\ \bibnamefont {Kim}},
  \bibinfo {author} {\bibfnamefont {J.}~\bibnamefont {Yu}}, \bibinfo {author}
  {\bibfnamefont {J.~H.}\ \bibnamefont {Han}}, \ and\ \bibinfo {author}
  {\bibfnamefont {C.}~\bibnamefont {Kim}},\ }\bibfield  {title} {\enquote
  {\bibinfo {title} {Orbital-angular-momentum based origin of {R}ashba-type
  surface band splitting},}\ }\href {\doibase 10.1103/PhysRevLett.107.156803}
  {\bibfield  {journal} {\bibinfo  {journal} {Phys. Rev. Lett.}\ }\textbf
  {\bibinfo {volume} {107}},\ \bibinfo {pages} {156803} (\bibinfo {year}
  {2011})}\BibitemShut {NoStop}%
\bibitem [{\citenamefont {Go}\ \emph {et~al.}(2017)\citenamefont {Go},
  \citenamefont {Hanke}, \citenamefont {Buhl}, \citenamefont {Freimuth},
  \citenamefont {Bihlmayer}, \citenamefont {Lee}, \citenamefont {Mokrousov},\
  and\ \citenamefont {Bl{\"u}gel}}]{Go2017}%
  \BibitemOpen
  \bibfield  {author} {\bibinfo {author} {\bibfnamefont {D.}~\bibnamefont
  {Go}}, \bibinfo {author} {\bibfnamefont {J.-P.}\ \bibnamefont {Hanke}},
  \bibinfo {author} {\bibfnamefont {P.~M.}\ \bibnamefont {Buhl}}, \bibinfo
  {author} {\bibfnamefont {F.}~\bibnamefont {Freimuth}}, \bibinfo {author}
  {\bibfnamefont {G.}~\bibnamefont {Bihlmayer}}, \bibinfo {author}
  {\bibfnamefont {H.-W.}\ \bibnamefont {Lee}}, \bibinfo {author} {\bibfnamefont
  {Y.}~\bibnamefont {Mokrousov}}, \ and\ \bibinfo {author} {\bibfnamefont
  {S.}~\bibnamefont {Bl{\"u}gel}},\ }\bibfield  {title} {\enquote {\bibinfo
  {title} {Toward surface orbitronics: giant orbital magnetism from the orbital
  {R}ashba effect at the surface of sp-metals},}\ }\href {\doibase
  10.1038/srep46742} {\bibfield  {journal} {\bibinfo  {journal} {Scientific
  Reports}\ }\textbf {\bibinfo {volume} {7}},\ \bibinfo {pages} {46742}
  (\bibinfo {year} {2017})}\BibitemShut {NoStop}%
\bibitem [{\citenamefont {Go}\ \emph {et~al.}(2021{\natexlab{b}})\citenamefont
  {Go}, \citenamefont {Jo}, \citenamefont {Gao}, \citenamefont {Ando},
  \citenamefont {Bl\"ugel}, \citenamefont {Lee},\ and\ \citenamefont
  {Mokrousov}}]{Go2021a}%
  \BibitemOpen
  \bibfield  {author} {\bibinfo {author} {\bibfnamefont {D.}~\bibnamefont
  {Go}}, \bibinfo {author} {\bibfnamefont {D.}~\bibnamefont {Jo}}, \bibinfo
  {author} {\bibfnamefont {T.}~\bibnamefont {Gao}}, \bibinfo {author}
  {\bibfnamefont {K.}~\bibnamefont {Ando}}, \bibinfo {author} {\bibfnamefont
  {S.}~\bibnamefont {Bl\"ugel}}, \bibinfo {author} {\bibfnamefont {H.-W.}\
  \bibnamefont {Lee}}, \ and\ \bibinfo {author} {\bibfnamefont
  {Y.}~\bibnamefont {Mokrousov}},\ }\bibfield  {title} {\enquote {\bibinfo
  {title} {Orbital {R}ashba effect in a surface-oxidized {C}u film},}\ }\href
  {\doibase 10.1103/PhysRevB.103.L121113} {\bibfield  {journal} {\bibinfo
  {journal} {Phys. Rev. B}\ }\textbf {\bibinfo {volume} {103}},\ \bibinfo
  {pages} {L121113} (\bibinfo {year} {2021}{\natexlab{b}})}\BibitemShut
  {NoStop}%
\bibitem [{\citenamefont {Calavalle}\ \emph {et~al.}(2022)\citenamefont
  {Calavalle}, \citenamefont {Su{\'a}rez-Rodr{\'i}guez}, \citenamefont
  {Mart{\'i}n-Garc{\'i}a}, \citenamefont {Johansson}, \citenamefont {Vaz},
  \citenamefont {Yang}, \citenamefont {Maznichenko}, \citenamefont {Ostanin},
  \citenamefont {Mateo-Alonso}, \citenamefont {Chuvilin}, \citenamefont
  {Mertig}, \citenamefont {Gobbi}, \citenamefont {Casanova},\ and\
  \citenamefont {Hueso}}]{Calavalle2022}%
  \BibitemOpen
  \bibfield  {author} {\bibinfo {author} {\bibfnamefont {F.}~\bibnamefont
  {Calavalle}}, \bibinfo {author} {\bibfnamefont {M.}~\bibnamefont
  {Su{\'a}rez-Rodr{\'i}guez}}, \bibinfo {author} {\bibfnamefont
  {B.}~\bibnamefont {Mart{\'i}n-Garc{\'i}a}}, \bibinfo {author} {\bibfnamefont
  {A.}~\bibnamefont {Johansson}}, \bibinfo {author} {\bibfnamefont {D.~C.}\
  \bibnamefont {Vaz}}, \bibinfo {author} {\bibfnamefont {H.}~\bibnamefont
  {Yang}}, \bibinfo {author} {\bibfnamefont {I.~V.}\ \bibnamefont
  {Maznichenko}}, \bibinfo {author} {\bibfnamefont {S.}~\bibnamefont
  {Ostanin}}, \bibinfo {author} {\bibfnamefont {A.}~\bibnamefont
  {Mateo-Alonso}}, \bibinfo {author} {\bibfnamefont {A.}~\bibnamefont
  {Chuvilin}}, \bibinfo {author} {\bibfnamefont {I.}~\bibnamefont {Mertig}},
  \bibinfo {author} {\bibfnamefont {M.}~\bibnamefont {Gobbi}}, \bibinfo
  {author} {\bibfnamefont {F.}~\bibnamefont {Casanova}}, \ and\ \bibinfo
  {author} {\bibfnamefont {L.~E.}\ \bibnamefont {Hueso}},\ }\bibfield  {title}
  {\enquote {\bibinfo {title} {Gate-tuneable and chirality-dependent
  charge-to-spin conversion in tellurium nanowires},}\ }\href {\doibase
  10.1038/s41563-022-01211-7} {\bibfield  {journal} {\bibinfo  {journal}
  {Nature Materials}\ }\textbf {\bibinfo {volume} {21}},\ \bibinfo {pages}
  {526--532} (\bibinfo {year} {2022})}\BibitemShut {NoStop}%
\bibitem [{\citenamefont {Roy}\ \emph {et~al.}(2022)\citenamefont {Roy},
  \citenamefont {Cerasoli}, \citenamefont {Jayaraj}, \citenamefont {Tenzin},
  \citenamefont {Nardelli},\ and\ \citenamefont
  {S{\l}awi{\'{n}}ska}}]{Roy2022}%
  \BibitemOpen
  \bibfield  {author} {\bibinfo {author} {\bibfnamefont {A.}~\bibnamefont
  {Roy}}, \bibinfo {author} {\bibfnamefont {F.~T.}\ \bibnamefont {Cerasoli}},
  \bibinfo {author} {\bibfnamefont {A.}~\bibnamefont {Jayaraj}}, \bibinfo
  {author} {\bibfnamefont {K.}~\bibnamefont {Tenzin}}, \bibinfo {author}
  {\bibfnamefont {M.~B.}\ \bibnamefont {Nardelli}}, \ and\ \bibinfo {author}
  {\bibfnamefont {J.}~\bibnamefont {S{\l}awi{\'{n}}ska}},\ }\bibfield  {title}
  {\enquote {\bibinfo {title} {Long-range current-induced spin accumulation in
  chiral crystals},}\ }\href {\doibase 10.1038/s41524-022-00931-3} {\bibfield
  {journal} {\bibinfo  {journal} {npj Computational Materials}\ }\textbf
  {\bibinfo {volume} {8}},\ \bibinfo {pages} {243} (\bibinfo {year}
  {2022})}\BibitemShut {NoStop}%
\bibitem [{\citenamefont {Ishito}\ \emph {et~al.}(2023)\citenamefont {Ishito},
  \citenamefont {Mao}, \citenamefont {Kousaka}, \citenamefont {Togawa},
  \citenamefont {Iwasaki}, \citenamefont {Zhang}, \citenamefont {Murakami},
  \citenamefont {Kishine},\ and\ \citenamefont {Satoh}}]{Ishito2023}%
  \BibitemOpen
  \bibfield  {author} {\bibinfo {author} {\bibfnamefont {K.}~\bibnamefont
  {Ishito}}, \bibinfo {author} {\bibfnamefont {H.}~\bibnamefont {Mao}},
  \bibinfo {author} {\bibfnamefont {Y.}~\bibnamefont {Kousaka}}, \bibinfo
  {author} {\bibfnamefont {Y.}~\bibnamefont {Togawa}}, \bibinfo {author}
  {\bibfnamefont {S.}~\bibnamefont {Iwasaki}}, \bibinfo {author} {\bibfnamefont
  {T.}~\bibnamefont {Zhang}}, \bibinfo {author} {\bibfnamefont
  {S.}~\bibnamefont {Murakami}}, \bibinfo {author} {\bibfnamefont {J.-i.}\
  \bibnamefont {Kishine}}, \ and\ \bibinfo {author} {\bibfnamefont
  {T.}~\bibnamefont {Satoh}},\ }\bibfield  {title} {\enquote {\bibinfo {title}
  {Truly chiral phonons in $\alpha$-{H}g{S}},}\ }\href {\doibase
  10.1038/s41567-022-01790-x} {\bibfield  {journal} {\bibinfo  {journal}
  {Nature Physics}\ }\textbf {\bibinfo {volume} {19}},\ \bibinfo {pages}
  {35--39} (\bibinfo {year} {2023})}\BibitemShut {NoStop}%
\bibitem [{\citenamefont {Ueda}\ \emph {et~al.}(2023)\citenamefont {Ueda},
  \citenamefont {Garc{\'i}a-Fern{\'a}ndez}, \citenamefont {Agrestini},
  \citenamefont {Romao}, \citenamefont {van~den Brink}, \citenamefont
  {Spaldin}, \citenamefont {Zhou},\ and\ \citenamefont {Staub}}]{Ueda2023}%
  \BibitemOpen
  \bibfield  {author} {\bibinfo {author} {\bibfnamefont {H.}~\bibnamefont
  {Ueda}}, \bibinfo {author} {\bibfnamefont {M.}~\bibnamefont
  {Garc{\'i}a-Fern{\'a}ndez}}, \bibinfo {author} {\bibfnamefont
  {S.}~\bibnamefont {Agrestini}}, \bibinfo {author} {\bibfnamefont {C.~P.}\
  \bibnamefont {Romao}}, \bibinfo {author} {\bibfnamefont {J.}~\bibnamefont
  {van~den Brink}}, \bibinfo {author} {\bibfnamefont {N.~A.}\ \bibnamefont
  {Spaldin}}, \bibinfo {author} {\bibfnamefont {K.-J.}\ \bibnamefont {Zhou}}, \
  and\ \bibinfo {author} {\bibfnamefont {U.}~\bibnamefont {Staub}},\ }\bibfield
   {title} {\enquote {\bibinfo {title} {Chiral phonons in quartz probed by
  x-rays},}\ }\href {\doibase 10.1038/s41586-023-06016-5} {\bibfield  {journal}
  {\bibinfo  {journal} {Nature}\ }\textbf {\bibinfo {volume} {618}},\ \bibinfo
  {pages} {946--950} (\bibinfo {year} {2023})}\BibitemShut {NoStop}%
\bibitem [{\citenamefont {Ou-Yang}\ \emph
  {et~al.}(2015{\natexlab{a}})\citenamefont {Ou-Yang}, \citenamefont {Shu},
  \citenamefont {Hu},\ and\ \citenamefont {Chou}}]{OuYang2015}%
  \BibitemOpen
  \bibfield  {author} {\bibinfo {author} {\bibfnamefont {T.}~\bibnamefont
  {Ou-Yang}}, \bibinfo {author} {\bibfnamefont {G.}~\bibnamefont {Shu}},
  \bibinfo {author} {\bibfnamefont {C.}~\bibnamefont {Hu}}, \ and\ \bibinfo
  {author} {\bibfnamefont {F.}~\bibnamefont {Chou}},\ }\bibfield  {title}
  {\enquote {\bibinfo {title} {Dynamic susceptibility study on the skyrmion
  phase stability of {F}e$_{0,7}${C}o$_{0.3}${S}i},}\ }\href {\doibase
  10.1063/1.4915934} {\bibfield  {journal} {\bibinfo  {journal} {J. Appl.
  Phys.}\ }\textbf {\bibinfo {volume} {117}},\ \bibinfo {pages} {123903}
  (\bibinfo {year} {2015}{\natexlab{a}})}\BibitemShut {NoStop}%
\bibitem [{\citenamefont {Ou-Yang}\ \emph
  {et~al.}(2015{\natexlab{b}})\citenamefont {Ou-Yang}, \citenamefont {Shu},
  \citenamefont {Hu},\ and\ \citenamefont {Chou}}]{OuYang2015a}%
  \BibitemOpen
  \bibfield  {author} {\bibinfo {author} {\bibfnamefont {T.}~\bibnamefont
  {Ou-Yang}}, \bibinfo {author} {\bibfnamefont {G.}~\bibnamefont {Shu}},
  \bibinfo {author} {\bibfnamefont {C.}~\bibnamefont {Hu}}, \ and\ \bibinfo
  {author} {\bibfnamefont {F.}~\bibnamefont {Chou}},\ }\bibfield  {title}
  {\enquote {\bibinfo {title} {Preparation of anomalous magnetoresistance and
  transport properties of itinerant ferromagnet {F}e$_{1-x}${C}o$_{x}${S}i},}\
  }\href {\doibase 10.1109/tmag.2015.2434882} {\bibfield  {journal} {\bibinfo
  {journal} {IEEE Trans. Magn.}\ }\textbf {\bibinfo {volume} {51}},\ \bibinfo
  {pages} {1700104} (\bibinfo {year} {2015}{\natexlab{b}})}\BibitemShut
  {NoStop}%
\bibitem [{\citenamefont {Castro}\ \emph {et~al.}(1997)\citenamefont {Castro},
  \citenamefont {Alvarez}, \citenamefont {D{\'a}vila}, \citenamefont
  {Asensio},\ and\ \citenamefont {Michel}}]{Castro1997}%
  \BibitemOpen
  \bibfield  {author} {\bibinfo {author} {\bibfnamefont {G.}~\bibnamefont
  {Castro}}, \bibinfo {author} {\bibfnamefont {J.}~\bibnamefont {Alvarez}},
  \bibinfo {author} {\bibfnamefont {M.}~\bibnamefont {D{\'a}vila}}, \bibinfo
  {author} {\bibfnamefont {M.}~\bibnamefont {Asensio}}, \ and\ \bibinfo
  {author} {\bibfnamefont {E.}~\bibnamefont {Michel}},\ }\bibfield  {title}
  {\enquote {\bibinfo {title} {Electronic band structure of
  $\epsilon$-{F}e{S}i(100)},}\ }\href
  {https://iopscience.iop.org/article/10.1088/0953-8984/9/8/017/meta}
  {\bibfield  {journal} {\bibinfo  {journal} {J. Phys.: Condens. Matter}\
  }\textbf {\bibinfo {volume} {9}},\ \bibinfo {pages} {1871} (\bibinfo {year}
  {1997})}\BibitemShut {NoStop}%
\bibitem [{\citenamefont {Klein}\ \emph {et~al.}(2009)\citenamefont {Klein},
  \citenamefont {Menzel}, \citenamefont {Doll}, \citenamefont {Neef},
  \citenamefont {Zur}, \citenamefont {Jursic}, \citenamefont {Schoenes},\ and\
  \citenamefont {Reinert}}]{Klein2009}%
  \BibitemOpen
  \bibfield  {author} {\bibinfo {author} {\bibfnamefont {M.}~\bibnamefont
  {Klein}}, \bibinfo {author} {\bibfnamefont {D.}~\bibnamefont {Menzel}},
  \bibinfo {author} {\bibfnamefont {K.}~\bibnamefont {Doll}}, \bibinfo {author}
  {\bibfnamefont {M.}~\bibnamefont {Neef}}, \bibinfo {author} {\bibfnamefont
  {D.}~\bibnamefont {Zur}}, \bibinfo {author} {\bibfnamefont {I.}~\bibnamefont
  {Jursic}}, \bibinfo {author} {\bibfnamefont {J.}~\bibnamefont {Schoenes}}, \
  and\ \bibinfo {author} {\bibfnamefont {F.}~\bibnamefont {Reinert}},\
  }\bibfield  {title} {\enquote {\bibinfo {title} {Photoemission spectroscopy
  across the semiconductor-to-metal transition in {F}e{S}i},}\ }\href {\doibase
  10.1088/1367-2630/11/2/023026} {\bibfield  {journal} {\bibinfo  {journal}
  {New J. Phys.}\ }\textbf {\bibinfo {volume} {11}},\ \bibinfo {pages} {023026}
  (\bibinfo {year} {2009})}\BibitemShut {NoStop}%
\bibitem [{\citenamefont {Kura}\ \emph {et~al.}(2008)\citenamefont {Kura},
  \citenamefont {Takano}, \citenamefont {Takeichi}, \citenamefont {Harasawa},
  \citenamefont {Okuda}, \citenamefont {Matsuda},\ and\ \citenamefont
  {Kakizaki}}]{Kura2008}%
  \BibitemOpen
  \bibfield  {author} {\bibinfo {author} {\bibfnamefont {K.}~\bibnamefont
  {Kura}}, \bibinfo {author} {\bibfnamefont {K.}~\bibnamefont {Takano}},
  \bibinfo {author} {\bibfnamefont {Y.}~\bibnamefont {Takeichi}}, \bibinfo
  {author} {\bibfnamefont {A.}~\bibnamefont {Harasawa}}, \bibinfo {author}
  {\bibfnamefont {T.}~\bibnamefont {Okuda}}, \bibinfo {author} {\bibfnamefont
  {I.}~\bibnamefont {Matsuda}}, \ and\ \bibinfo {author} {\bibfnamefont
  {A.}~\bibnamefont {Kakizaki}},\ }\bibfield  {title} {\enquote {\bibinfo
  {title} {Weak electron correlation effects observed in angle-resolved
  photoemission spectra of {M}n{S}i (100)},}\ }\href {\doibase
  10.1143/jpsj.77.024709} {\bibfield  {journal} {\bibinfo  {journal} {J. Phys.
  Soc. Jpn.}\ }\textbf {\bibinfo {volume} {77}},\ \bibinfo {pages}
  {024709--024709} (\bibinfo {year} {2008})}\BibitemShut {NoStop}%
\bibitem [{\citenamefont {Tusche}\ \emph
  {et~al.}(2019{\natexlab{b}})\citenamefont {Tusche}, \citenamefont {Chen},
  \citenamefont {Schneider},\ and\ \citenamefont {Kirschner}}]{tus19}%
  \BibitemOpen
  \bibfield  {author} {\bibinfo {author} {\bibfnamefont {C.}~\bibnamefont
  {Tusche}}, \bibinfo {author} {\bibfnamefont {Y.-J.}\ \bibnamefont {Chen}},
  \bibinfo {author} {\bibfnamefont {C.~M.}\ \bibnamefont {Schneider}}, \ and\
  \bibinfo {author} {\bibfnamefont {J.}~\bibnamefont {Kirschner}},\ }\bibfield
  {title} {\enquote {\bibinfo {title} {Imaging properties of hemispherical
  electrostatic energy analyzers for high resolution momentum microscopy},}\
  }\href {\doibase 10.1016/j.ultramic.2019.112815} {\bibfield  {journal}
  {\bibinfo  {journal} {Ultramicroscopy}\ }\textbf {\bibinfo {volume} {206}},\
  \bibinfo {pages} {112815} (\bibinfo {year} {2019}{\natexlab{b}})}\BibitemShut
  {NoStop}%
\bibitem [{\citenamefont {Wortmann}\ \emph {et~al.}(2023)\citenamefont
  {Wortmann}, \citenamefont {Michalicek}, \citenamefont {Baadji}, \citenamefont
  {Betzinger}, \citenamefont {Bihlmayer}, \citenamefont {Br\"oder},
  \citenamefont {Burnus}, \citenamefont {Enkovaara}, \citenamefont {Freimuth},
  \citenamefont {Friedrich}, \citenamefont {Gerhorst}, \citenamefont
  {Granberg~Cauchi}, \citenamefont {Grytsiuk}, \citenamefont {Hanke},
  \citenamefont {Hanke}, \citenamefont {Heide}, \citenamefont {Heinze},
  \citenamefont {Hilgers}, \citenamefont {Janssen}, \citenamefont
  {Kl\"uppelberg}, \citenamefont {Kovacik}, \citenamefont {Kurz}, \citenamefont
  {Lezaic}, \citenamefont {Madsen}, \citenamefont {Mokrousov}, \citenamefont
  {Neukirchen}, \citenamefont {Redies}, \citenamefont {Rost}, \citenamefont
  {Schlipf}, \citenamefont {Schindlmayr}, \citenamefont {Winkelmann},\ and\
  \citenamefont {Bl\"ugel}}]{fleur}%
  \BibitemOpen
  \bibfield  {author} {\bibinfo {author} {\bibfnamefont {D.}~\bibnamefont
  {Wortmann}}, \bibinfo {author} {\bibfnamefont {G.}~\bibnamefont
  {Michalicek}}, \bibinfo {author} {\bibfnamefont {N.}~\bibnamefont {Baadji}},
  \bibinfo {author} {\bibfnamefont {M.}~\bibnamefont {Betzinger}}, \bibinfo
  {author} {\bibfnamefont {G.}~\bibnamefont {Bihlmayer}}, \bibinfo {author}
  {\bibfnamefont {J.}~\bibnamefont {Br\"oder}}, \bibinfo {author}
  {\bibfnamefont {T.}~\bibnamefont {Burnus}}, \bibinfo {author} {\bibfnamefont
  {J.}~\bibnamefont {Enkovaara}}, \bibinfo {author} {\bibfnamefont
  {F.}~\bibnamefont {Freimuth}}, \bibinfo {author} {\bibfnamefont
  {C.}~\bibnamefont {Friedrich}}, \bibinfo {author} {\bibfnamefont {C.-R.}\
  \bibnamefont {Gerhorst}}, \bibinfo {author} {\bibfnamefont {S.}~\bibnamefont
  {Granberg~Cauchi}}, \bibinfo {author} {\bibfnamefont {U.}~\bibnamefont
  {Grytsiuk}}, \bibinfo {author} {\bibfnamefont {A.}~\bibnamefont {Hanke}},
  \bibinfo {author} {\bibfnamefont {J.-P.}\ \bibnamefont {Hanke}}, \bibinfo
  {author} {\bibfnamefont {M.}~\bibnamefont {Heide}}, \bibinfo {author}
  {\bibfnamefont {S.}~\bibnamefont {Heinze}}, \bibinfo {author} {\bibfnamefont
  {R.}~\bibnamefont {Hilgers}}, \bibinfo {author} {\bibfnamefont
  {H.}~\bibnamefont {Janssen}}, \bibinfo {author} {\bibfnamefont {D.~A.}\
  \bibnamefont {Kl\"uppelberg}}, \bibinfo {author} {\bibfnamefont
  {R.}~\bibnamefont {Kovacik}}, \bibinfo {author} {\bibfnamefont
  {P.}~\bibnamefont {Kurz}}, \bibinfo {author} {\bibfnamefont {M.}~\bibnamefont
  {Lezaic}}, \bibinfo {author} {\bibfnamefont {G.~K.~H.}\ \bibnamefont
  {Madsen}}, \bibinfo {author} {\bibfnamefont {Y.}~\bibnamefont {Mokrousov}},
  \bibinfo {author} {\bibfnamefont {A.}~\bibnamefont {Neukirchen}}, \bibinfo
  {author} {\bibfnamefont {M.}~\bibnamefont {Redies}}, \bibinfo {author}
  {\bibfnamefont {S.}~\bibnamefont {Rost}}, \bibinfo {author} {\bibfnamefont
  {M.}~\bibnamefont {Schlipf}}, \bibinfo {author} {\bibfnamefont
  {A.}~\bibnamefont {Schindlmayr}}, \bibinfo {author} {\bibfnamefont
  {M.}~\bibnamefont {Winkelmann}}, \ and\ \bibinfo {author} {\bibfnamefont
  {S.}~\bibnamefont {Bl\"ugel}},\ }\href {\doibase 10.5281/zenodo.7576163}
  {\enquote {\bibinfo {title} {{FLEUR}},}\ }\bibinfo {howpublished} {Zenodo}
  (\bibinfo {year} {2023})\BibitemShut {NoStop}%
\bibitem [{\citenamefont {Wimmer}\ \emph {et~al.}(1981)\citenamefont {Wimmer},
  \citenamefont {Krakauer}, \citenamefont {Weinert},\ and\ \citenamefont
  {Freeman}}]{Wimmer1981}%
  \BibitemOpen
  \bibfield  {author} {\bibinfo {author} {\bibfnamefont {E.}~\bibnamefont
  {Wimmer}}, \bibinfo {author} {\bibfnamefont {H.}~\bibnamefont {Krakauer}},
  \bibinfo {author} {\bibfnamefont {M.}~\bibnamefont {Weinert}}, \ and\
  \bibinfo {author} {\bibfnamefont {A.~J.}\ \bibnamefont {Freeman}},\
  }\bibfield  {title} {\enquote {\bibinfo {title} {{Full-potential
  self-consistent linearized-augmented-plane-wave method for calculating the
  electronic structure of molecules and surfaces: ${\mathrm{O}}_{2}$
  molecule}},}\ }\href {\doibase 10.1103/PhysRevB.24.864} {\bibfield  {journal}
  {\bibinfo  {journal} {Phys. Rev. B}\ }\textbf {\bibinfo {volume} {24}},\
  \bibinfo {pages} {864--875} (\bibinfo {year} {1981})}\BibitemShut {NoStop}%
\bibitem [{\citenamefont {Perdew}, \citenamefont {Burke},\ and\ \citenamefont
  {Ernzerhof}(1996)}]{Perdew1996}%
  \BibitemOpen
  \bibfield  {author} {\bibinfo {author} {\bibfnamefont {J.~P.}\ \bibnamefont
  {Perdew}}, \bibinfo {author} {\bibfnamefont {K.}~\bibnamefont {Burke}}, \
  and\ \bibinfo {author} {\bibfnamefont {M.}~\bibnamefont {Ernzerhof}},\
  }\bibfield  {title} {\enquote {\bibinfo {title} {{Generalized Gradient
  Approximation Made Simple}},}\ }\href {\doibase 10.1103/PhysRevLett.77.3865}
  {\bibfield  {journal} {\bibinfo  {journal} {Phys. Rev. Lett.}\ }\textbf
  {\bibinfo {volume} {77}},\ \bibinfo {pages} {3865--3868} (\bibinfo {year}
  {1996})}\BibitemShut {NoStop}%
\bibitem [{\citenamefont {Pizzi}\ \emph {et~al.}(2020)\citenamefont {Pizzi},
  \citenamefont {Vitale}, \citenamefont {Arita}, \citenamefont {Bl\"ugel},
  \citenamefont {Freimuth}, \citenamefont {G{\'{e}}ranton}, \citenamefont
  {Gibertini}, \citenamefont {Gresch}, \citenamefont {Johnson}, \citenamefont
  {Koretsune}, \citenamefont {Iba{\~{n}}ez-Azpiroz}, \citenamefont {Lee},
  \citenamefont {Lihm}, \citenamefont {Marchand}, \citenamefont {Marrazzo},
  \citenamefont {Mokrousov}, \citenamefont {Mustafa}, \citenamefont {Nohara},
  \citenamefont {Nomura}, \citenamefont {Paulatto}, \citenamefont
  {Ponc{\'{e}}}, \citenamefont {Ponweiser}, \citenamefont {Qiao}, \citenamefont
  {Th\"ole}, \citenamefont {Tsirkin}, \citenamefont {Wierzbowska},
  \citenamefont {Marzari}, \citenamefont {Vanderbilt}, \citenamefont {Souza},
  \citenamefont {Mostofi},\ and\ \citenamefont {Yates}}]{Pizzi2020}%
  \BibitemOpen
  \bibfield  {author} {\bibinfo {author} {\bibfnamefont {G.}~\bibnamefont
  {Pizzi}}, \bibinfo {author} {\bibfnamefont {V.}~\bibnamefont {Vitale}},
  \bibinfo {author} {\bibfnamefont {R.}~\bibnamefont {Arita}}, \bibinfo
  {author} {\bibfnamefont {S.}~\bibnamefont {Bl\"ugel}}, \bibinfo {author}
  {\bibfnamefont {F.}~\bibnamefont {Freimuth}}, \bibinfo {author}
  {\bibfnamefont {G.}~\bibnamefont {G{\'{e}}ranton}}, \bibinfo {author}
  {\bibfnamefont {M.}~\bibnamefont {Gibertini}}, \bibinfo {author}
  {\bibfnamefont {D.}~\bibnamefont {Gresch}}, \bibinfo {author} {\bibfnamefont
  {C.}~\bibnamefont {Johnson}}, \bibinfo {author} {\bibfnamefont
  {T.}~\bibnamefont {Koretsune}}, \bibinfo {author} {\bibfnamefont
  {J.}~\bibnamefont {Iba{\~{n}}ez-Azpiroz}}, \bibinfo {author} {\bibfnamefont
  {H.}~\bibnamefont {Lee}}, \bibinfo {author} {\bibfnamefont {J.-M.}\
  \bibnamefont {Lihm}}, \bibinfo {author} {\bibfnamefont {D.}~\bibnamefont
  {Marchand}}, \bibinfo {author} {\bibfnamefont {A.}~\bibnamefont {Marrazzo}},
  \bibinfo {author} {\bibfnamefont {Y.}~\bibnamefont {Mokrousov}}, \bibinfo
  {author} {\bibfnamefont {J.~I.}\ \bibnamefont {Mustafa}}, \bibinfo {author}
  {\bibfnamefont {Y.}~\bibnamefont {Nohara}}, \bibinfo {author} {\bibfnamefont
  {Y.}~\bibnamefont {Nomura}}, \bibinfo {author} {\bibfnamefont
  {L.}~\bibnamefont {Paulatto}}, \bibinfo {author} {\bibfnamefont
  {S.}~\bibnamefont {Ponc{\'{e}}}}, \bibinfo {author} {\bibfnamefont
  {T.}~\bibnamefont {Ponweiser}}, \bibinfo {author} {\bibfnamefont
  {J.}~\bibnamefont {Qiao}}, \bibinfo {author} {\bibfnamefont {F.}~\bibnamefont
  {Th\"ole}}, \bibinfo {author} {\bibfnamefont {S.~S.}\ \bibnamefont
  {Tsirkin}}, \bibinfo {author} {\bibfnamefont {M.}~\bibnamefont
  {Wierzbowska}}, \bibinfo {author} {\bibfnamefont {N.}~\bibnamefont
  {Marzari}}, \bibinfo {author} {\bibfnamefont {D.}~\bibnamefont {Vanderbilt}},
  \bibinfo {author} {\bibfnamefont {I.}~\bibnamefont {Souza}}, \bibinfo
  {author} {\bibfnamefont {A.~A.}\ \bibnamefont {Mostofi}}, \ and\ \bibinfo
  {author} {\bibfnamefont {J.~R.}\ \bibnamefont {Yates}},\ }\bibfield  {title}
  {\enquote {\bibinfo {title} {Wannier90 as a community code: new features and
  applications},}\ }\href {\doibase 10.1088/1361-648x/ab51ff} {\bibfield
  {journal} {\bibinfo  {journal} {Journal of Physics: Condensed Matter}\
  }\textbf {\bibinfo {volume} {32}},\ \bibinfo {pages} {165902} (\bibinfo
  {year} {2020})}\BibitemShut {NoStop}%
\bibitem [{\citenamefont {Freimuth}\ \emph {et~al.}(2008)\citenamefont
  {Freimuth}, \citenamefont {Mokrousov}, \citenamefont {Wortmann},
  \citenamefont {Heinze},\ and\ \citenamefont {Bl\"ugel}}]{Freimuth2008}%
  \BibitemOpen
  \bibfield  {author} {\bibinfo {author} {\bibfnamefont {F.}~\bibnamefont
  {Freimuth}}, \bibinfo {author} {\bibfnamefont {Y.}~\bibnamefont {Mokrousov}},
  \bibinfo {author} {\bibfnamefont {D.}~\bibnamefont {Wortmann}}, \bibinfo
  {author} {\bibfnamefont {S.}~\bibnamefont {Heinze}}, \ and\ \bibinfo {author}
  {\bibfnamefont {S.}~\bibnamefont {Bl\"ugel}},\ }\bibfield  {title} {\enquote
  {\bibinfo {title} {{Maximally localized Wannier functions within the FLAPW
  formalism}},}\ }\href {\doibase 10.1103/PhysRevB.78.035120} {\bibfield
  {journal} {\bibinfo  {journal} {Phys. Rev. B}\ }\textbf {\bibinfo {volume}
  {78}},\ \bibinfo {pages} {035120} (\bibinfo {year} {2008})}\BibitemShut
  {NoStop}%
\bibitem [{\citenamefont {Hanke}\ \emph {et~al.}(2016)\citenamefont {Hanke},
  \citenamefont {Freimuth}, \citenamefont {Nandy}, \citenamefont {Zhang},
  \citenamefont {Bl\"ugel},\ and\ \citenamefont {Mokrousov}}]{Hanke2016}%
  \BibitemOpen
  \bibfield  {author} {\bibinfo {author} {\bibfnamefont {J.-P.}\ \bibnamefont
  {Hanke}}, \bibinfo {author} {\bibfnamefont {F.}~\bibnamefont {Freimuth}},
  \bibinfo {author} {\bibfnamefont {A.~K.}\ \bibnamefont {Nandy}}, \bibinfo
  {author} {\bibfnamefont {H.}~\bibnamefont {Zhang}}, \bibinfo {author}
  {\bibfnamefont {S.}~\bibnamefont {Bl\"ugel}}, \ and\ \bibinfo {author}
  {\bibfnamefont {Y.}~\bibnamefont {Mokrousov}},\ }\bibfield  {title} {\enquote
  {\bibinfo {title} {{Role of Berry phase theory for describing orbital
  magnetism: From magnetic heterostructures to topological orbital
  ferromagnets}},}\ }\href {\doibase 10.1103/PhysRevB.94.121114} {\bibfield
  {journal} {\bibinfo  {journal} {Phys. Rev. B}\ }\textbf {\bibinfo {volume}
  {94}},\ \bibinfo {pages} {121114(R)} (\bibinfo {year} {2016})}\BibitemShut
  {NoStop}%
\bibitem [{\citenamefont {Sancho}\ \emph {et~al.}(1985)\citenamefont {Sancho},
  \citenamefont {Sancho}, \citenamefont {Sancho},\ and\ \citenamefont
  {Rubio}}]{Sancho1985}%
  \BibitemOpen
  \bibfield  {author} {\bibinfo {author} {\bibfnamefont {M.~P.~L.}\
  \bibnamefont {Sancho}}, \bibinfo {author} {\bibfnamefont {J.~M.~L.}\
  \bibnamefont {Sancho}}, \bibinfo {author} {\bibfnamefont {J.~M.~L.}\
  \bibnamefont {Sancho}}, \ and\ \bibinfo {author} {\bibfnamefont
  {J.}~\bibnamefont {Rubio}},\ }\bibfield  {title} {\enquote {\bibinfo {title}
  {{Highly convergent schemes for the calculation of bulk and surface Green
  functions}},}\ }\href {\doibase 10.1088/0305-4608/15/4/009} {\bibfield
  {journal} {\bibinfo  {journal} {Journal of Physics F: Metal Physics}\
  }\textbf {\bibinfo {volume} {15}},\ \bibinfo {pages} {851} (\bibinfo {year}
  {1985})}\BibitemShut {NoStop}%
\end{thebibliography}%

\end{document}